\documentclass[twocolumn,amsmath,amssymb,pra]{revtex4-2}
\usepackage{graphicx}
\usepackage{amsmath,amsfonts,amssymb,amsthm}
\usepackage{fancyhdr}
\usepackage{mathrsfs}
\usepackage{epstopdf}
\usepackage{setspace}
\usepackage{bm}
\usepackage[colorlinks]{hyperref}
\usepackage{xcolor}
\usepackage{dcolumn}

\usepackage{tikz}
\usetikzlibrary{quantikz}
\graphicspath{{figures/}}

\begin{document}

\title{Demonstration of system-bath physics on a gate-based quantum computer}
\author{Pascal Stadler}
\author{Matteo Lodi}
\author{Andisheh Khedri}
\author{Rolando Reiner}
\author{Kirsten Bark}
\author{Nicolas Vogt}
\author{Michael Marthaler}
\author{Juha Leppäkangas}
\affiliation{HQS Quantum Simulations GmbH, Rintheimer Straße 23, 76131 Karlsruhe, Germany}


\begin{abstract}
Algorithmic cooling can be used to find correlated states of many-body quantum systems. 
It is based on quantum circuits that perform nonunitary operations, whose implementation can be challenging on near-term quantum computers.
In this work we develop a method that uses inherent qubit noise to implement nonunitary operations and algorithmic cooling.
In our approach, qubit decay during quantum computation is used to simulate dissipation of auxiliary-spin bath,
which cools down a simulated system towards its ground state.
We test the algorithm on IBM-Q devices and demonstrate the relaxation of system spins to ferromagnetic and antiferromagnetic ordering,
controlled by the definition of the system Hamiltonian. The ordering is stable as long as the algorithm is run.
We are able to perform cooling and state stabilization for global systems of up to three system spins and four auxiliary spins.
Our work paves the way for useful quantum simulations of many-body quantum systems on near-term quantum computers.
\end{abstract}


\maketitle

\section{Introduction}
\label{sec:introduction}

The quantum computers of today grapple with significant challenges posed by qubit noise and gate errors.
Therefore, prior to achieving full quantum error correction, a crucial hurdle lies in discovering quantum algorithms that can effectively operate in the presence of noise.
This involves either mitigating the noise or harnessing it as a resource.
One intriguing approach involves mapping the effects of noise during a quantum mechanical time evolution onto processes within  open quantum systems~\cite{Lloyd1996, Tseng2010, Fratus2022}. 
It can be argued that this should lead to the possibility to study almost any system in nature, since the very essence of nature is rooted in open systems.
Notably, the study of finite or zero-temperature environments, often referred to as ``baths'', holds particular interest.
Rather than randomizing the system, these baths seek to drive the system towards a new ordering.
Many fundamental models of molecules and materials rely on such system-bath effects,
including density-matrix embedding~\cite{Wouters2016, Sun_2020}, dynamical mean-field theory~\cite{Rozenberg1996}, and spin-boson theories~\cite{Spin_Boson_Rev, Ishizaki2012, Huelga_2013}.

In addition, the exploration of open and thermally distributed quantum systems could pave the way for advancements in quantum machine learning~\cite{Olivera2023}.
Many models of quantum machine learning rely on achieving a thermal distribution of a spin system and explicitly require some type of bath.
The typical examples here are quantum Boltzmann machines and quantum reservoir computing.
In a quantum Boltzmann machine the general goal is to optimize the parameters of a system of coupled spins in a thermal Boltzmann distribution~\cite{Moro2023}.
For quantum reservoir computing it has been explicitly proposed that the dissipative dynamics needed in reservoir computing can be achieved by considering the effects of natural noise in qubits~\cite{Suzuki2022}.

The key ingredient of algorithms simulating interaction with finite- or zero-temperature baths
(see for example Refs.~\cite{Barreiro2011, Wang2011, Raghunandan_2020, Leppakangas2023, Mi2023, Matthies2024, Puente2024, Marti2024, Lin2024, Lloyd2024})
is the implementation of nonunitary time-evolution with a dependency on the sign of the energy change.
This directionality has been realized with the help of additional qubits and specifically engineered qubit reset gates~\cite{Barreiro2011, Han2021, Rost2021, Mi2023}. 
In our work, we successfully implement the needed nonunitary operations via utilization of inherent noise during the quantum computation.
Alternative methods include coupling to a large set of fresh (in advance initialized) qubits~\cite{Cattaneo2023}. 
Open-system models can also be solved by representing the density matrix of the system
as a state vector using doubled amount of qubits, 
and combining this with variational algorithms~\cite{Yoshioka2020} or repeated state tomography~\cite{Kamakari2022}. 

It should be noted that, throughout the paper, we refer to ``spins'' as the degrees of freedom of the effective system that is simulated
by a quantum computer, and refer to ``qubits'' to indicate the actual quantum computer. The time evolution of the spins is described by a 
a master equation which is not explicitly time-dependent. The quantum computer is going to time-evolve an initial state using an algorithm
that mimics the master equation, while in fact rapidly applying various gates. Each spin is directly mapped to one qubit. However,
what is important for our work is that the noise acting on a qubit can be transformed substantially through the application of gates when considered
from the perspective of the spins. We discussed these effects at length in Ref.~\cite{Fratus2022}. To a certain extent, we will make use of these transformations
by creating an effective environment at infinite temperature. However, we will try to balance this with avoiding these transformations where possible, in order to achieve the directionality
provided by qubit damping.

In this work, we follow closely the proposal of Ref.~\cite{Leppakangas2023} and divide the available qubits into system and bath qubits,
which represent the system we want to cool and a bath of auxiliary spins.
The used algorithm is based on utilizing bath-qubit damping that occurs during the quantum computation.
In this algorithm,
the interactions within the system are simulated by standard Trotterization of the unitary time-evolution operator,
whereas the energy-absorbing bath is simulated by combining the Trotterized time evolution of the auxiliary spins and the inherent damping of qubits representing them.
The steady state of the system spins can be made to prefer populating lower-energy eigenstates, which in turn
can then be altered by choosing different system Hamiltonians.
We test the algorithm on IBM-Q quantum computers which are available via cloud access.
We demonstrate the feasibility of this approach 
through the simulated relaxation of a system of interacting spins towards ferromagnetic or antiferromagnetic ordering.
We are able to simulate global systems of up to three system spins and four auxiliary spins.
We show that the steady-state correlations are stable under the inherent noise and last as long as the algorithm is being executed.
Our work presents, to the best of our knowledge, the first demonstration of algorithmic cooling on a publicly available quantum computer.

The outline of this paper is as follows.
In Section~\ref{sec:system_bath},
we describe the open quantum system model we consider in this work. 
In Section~\ref{sec:implementation_on_hardware},
we describe how it is implemented on a digital quantum computer using inherent noise.
In Section~\ref{sec:results}, we present our results on running the algorithm on the IBM-Q devices.
We conclude in Section~\ref{sec:conclusion}.
In Appendix~\ref{sec:spectrum_sketch},
we discuss the modeled system-bath physics and the fundamental limitations of the cooling efficiency in terms of a bath spectral function.
In Appendix~\ref{sec:NoiseAnalysis}, we present our noise tomography performed on the IBM quantum computers.
In Appendix~\ref{sec:appCircuits}, we describe in more detail the circuits submitted to the IBM-Q devices.
Finally, in Appendix~\ref{sec:appNumericalModel},
we describe our numerical model of noisy quantum computing,
used to fit and better understand the obtained results.
The submitted jobs and the numerical simulations are also available online at Ref.~\cite{qoqo_files_online}
in a format compatible with the qoqo toolkit~\cite{qoqo}.


\section{Simulated open quantum system}
\label{sec:system_bath}

\begin{figure}
\begin{center}
\includegraphics[width=1\columnwidth]{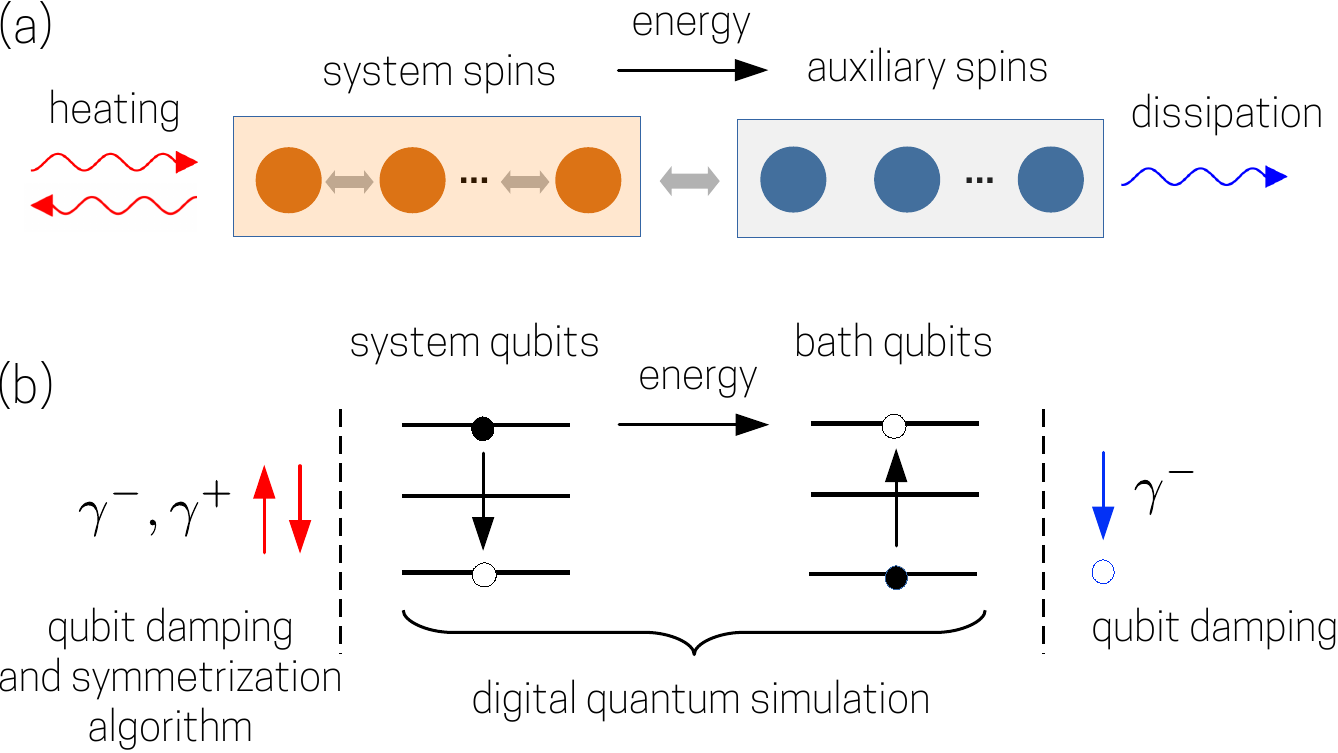}
\end{center}
\caption{
(a) We consider an open-system model where a quantum system of spins is cooled towards its ground state via coupling to auxiliary spins with dissipation.
Cooling is a result of resonant interaction between the system and the auxiliary spins with damping. 
(b) On the gate-based quantum computer with qubit damping, the energy-level structure and the time evolution of the global system is implemented 
digitally, whereas the dissipation of the auxiliary spins is a result of bath-qubit damping.
The simulated system is also connected to an additional environment causing random transitions between its energy levels and thereby heating.
The origin of this is system-qubit damping, which is seen as system-spin heating
when the introduced time-propagation algorithm is used (see Sec.~\ref{sec:implementation_on_hardware}).
}\label{fig:cooling}
\end{figure}

The open system considered is shown in Fig.~\ref{fig:cooling}(a).
It consists of a system of interacting spins coupled to an auxiliary-spin bath.
The auxiliary spins couple resonantly to the system,
so that auxiliary spin energy splittings exist close to the relevant energy-level differences in the system. 
The auxiliary spins are also subjected to damping, which dissipates energy from the bath,
and, through the resonant interaction, also provides cooling for the system.
The system is also connected to an additional environment,
causing random transitions between its energy levels and thereby heating.
We call the additional environment ``background'' since it directly acts on the system and is not mediated through the auxiliary-spin bath.
Its interpretation in terms of a bath spectral function is given in Appendix~\ref{sec:spectrum_sketch}.

The Hamiltonian that describes our total (global) system  can be written in the form
\begin{align}\label{eq:total_system_Hamiltonian}
\hat H &= \hat H_\textrm{S} + \hat H_\textrm{B} + \hat H_\textrm{C} \, ,
\end{align}
where $\hat H_\textrm{S}$ describes the system spins, $\hat H_\textrm{B}$ the auxiliary-spin bath,
and $\hat H_\textrm{C}$ the coupling between these two. The decoherence of systems spins and the auxiliary-spin bath
is included within the Lindblad master-equation formulation, defined below.

In this work, we consider spin systems with longitudinal nearest-neighbor coupling~$g_{ii'}$ 
and energy splittings $\epsilon_{\textrm{s},i}$ ($\hbar=1$),
\begin{align}
\hat H_\textrm{S} &= -\sum_i \frac{\epsilon_{\textrm{s},i}}{2} \hat \sigma_z^{\textrm{s},i} + \sum_{<ii'>}g_{ii'} \hat\sigma_z^{\textrm{s},i} \hat\sigma_z^{\textrm{s},i'} \, . \label{eq:system_Hamiltonian}
\end{align}
It should be noted that  the approach is not restricted to this choice
of the system Hamiltonian.

The auxiliary spins are chosen to be non-interacting
\begin{align}
\hat H_\textrm{B} &= -\sum_j \frac{\epsilon_{\textrm{b},j}}{2} \hat \sigma_z^{\textrm{b},j} \, . \label{eq:bath_Hamiltonian}
\end{align}
This choice is motivated by the need to keep the bath Hamiltonian simple
and thereby avoid unwanted transformations of the decoherence. Namely,
we want to ensure that qubit damping maps to the damping of the auxiliary-spins in the simulated model (see Sec.~\ref{sec:noise_mapping}).
The coupling between the system and bath is chosen to be
\begin{align}
\hat H_\textrm{C} &= \sum_{ij} v_{ij} \hat \sigma_x^{\textrm{s},i}\hat \sigma_x^{\textrm{b},j}  \, . \label{eq:coupling_Hamiltonian}
\end{align}
The coupling induces transitions between eigenstates of the system, $[\hat{H}_\textrm{S}, \hat{H}_\textrm{C}]\neq 0$,
and simultaneously excites the bath from its ground state.

The noise processes are included by introducing a Lindblad master-equation of the form
\begin{align}\label{eq:Lindbladian}
\dot {\hat \rho} &= \textrm{i}[\hat \rho, \hat H] + {\cal L}_\textrm{d}[\hat\rho] + {\cal L}_\textrm{h}[\hat\rho]  \, .
\end{align}
where~$\hat\rho$ is the density matrix of the global system, i.e., the system and auxiliary spins.
The auxiliary-spin damping channel is represented by Lindbladian~${\cal L}_\textrm{d}$ and the system-spin heating channel
by Lindbladian~${\cal L}_\textrm{h}$, which have the form
\begin{align}
{\cal L}_\textrm{d}[\hat\rho] &= \sum_{j}\gamma^-_j\left( \hat \sigma_-^{\textrm{b},j} \hat \rho \hat \sigma_+^{\textrm{b},j} - \frac{1}{2}\left\{\hat \sigma_+^{\textrm{b},j} \sigma_-^{\textrm{b},j} , \hat\rho \right\} \right) \label{eq:Lindbladian2}\\
{\cal L}_\textrm{h}[\hat\rho] &= \sum_{i}\gamma^x_{i}\left( \hat \sigma_x^{\textrm{s},i} \hat \rho \hat \sigma_x^{\textrm{s},i} - \hat\rho \right) + \sum_{i}\gamma^y_{i}\left( \hat \sigma_y^{\textrm{s},i} \hat \rho \hat \sigma_y^{\textrm{s},i} - \hat\rho \right) \nonumber\\
&+ \sum_{i}\gamma^z_{i}\left( \hat \sigma_z^{\textrm{s},i} \hat \rho \hat \sigma_z^{\textrm{s},i} - \hat\rho \right)  \label{eq:Lindbladian3}\, .
\end{align}
In particular, the system spins are subjected to Pauli noise, 
which can be interpret to be an effect of coupling of an infinite-temperature background to the system spins. 
We already note that in the quantum simulation of spins,
all noise channels can originate in qubit damping during execution of the quantum algorithm. 
In particular, damping of system qubits maps to the system-spin heating via the time propagation algorithm 
introduced in Sec.~\ref{sec:implementation_on_hardware}. 
If this is the only noise acting on system qubits, we have $\gamma^x_{i}=\gamma^y_{i}$ and $\gamma^z_{i}=0$. 
In Eq.~(\ref{eq:Lindbladian3}), we also allow for a finite z-component,
which can originate in dephasing or depolarising noise of system qubits. 
It should be noted that such noise can also be tolerated in the bath, 
but it reduces the efficiency of the algorithmic cooling. 

We can still formally trace out the auxiliary bath degrees of freedom and write down  
a Pauli master equation for eigenstate populations of the system,
\begin{align}
\dot\rho_{n} &=  \sum_k \left(\Gamma_{n\leftarrow k} \rho_{k} - \Gamma_{k\leftarrow n} \rho_{n}\right) \, ,
\end{align}
where~$\rho_n \equiv [\hat\rho_\textrm{s}]_{nn}$ is the population of the system eigenstate~$n$. 
This equation is valid in the limit of weak system-bath couplings compared to bath dissipation rates
and small transition rates~$\Gamma$ compared to energy-level differences.
The minimal goal of algorithmic cooling is then to achieve a situation where $\Gamma_{n \leftarrow k} > \Gamma_{k \leftarrow n}$ for system eigenenergies~$E_n < E_k$.
The form of the transition rates as a function of the bath and coupling parameters is discussed more detailed in Appendix~\ref{sec:spectrum_sketch}.


\section{Implementation on a noisy quantum computer}
\label{sec:implementation_on_hardware}
The goal is now to establish a quantum algorithm that time-propagates the simulated global system on a quantum computer,
as described by Eqs.~(\ref{eq:Lindbladian}-\ref{eq:Lindbladian3}).
As visualized in Fig.~\ref{fig:cooling}(b),
in our ideal realization, 
the energy-level structure and the time evolution of the global system will be implemented 
digitally, while the simulated auxiliary-spin dissipation originates in bath-qubit damping.
The background, which causes heating of the system, originates in system-qubit damping,
which has been tailored to look like heating using the system-noise symmetrization algorithm detailed below.
Of course, also other forms of qubit noise (and thereby also spin noise) may appear in a practical realization on hardware.

\subsection{Coherent time evolution}\label{sec:time_propagation}
In our realization, 
each simulated spin is directly mapped to one qubit on a quantum computer~\cite{Leppakangas2023}.
The coherent time evolution of the spins, i.e., the commutator with $\hat H$,
is implemented via Trotterization of the time-evolution operator~\cite{Childs2021}
\begin{align}\label{eq:time_propagation}
e^{-\textrm{i} \hat H t} \approx \left[ \Pi_{k} e^{-\textrm{i} \hat H_k \tau} \right]^m  \, ,
\end{align}
where the total simulated time $t$ is divided into $m$ Trotter time-steps~$\tau$, so we have $t=m\tau$.
The chosen partial Hamiltonians $\hat H_k$ satisfy $\hat H=\sum_k \hat H_k$ and correspond to spin splittings or interaction terms between two spins.
The unitary operations $e^{-\textrm{i} \hat H_k \tau}$ are implemented
using the available unitary gates~$\hat U$ on the quantum computer,
\begin{align}\label{eq:gate_decomposition}
e^{-\textrm{i} \hat H_k \tau} = \Pi_l \hat U_{k l}  \, ,
\end{align}
and are chosen such that the simultaneously implemented incoherent time evolution is in the correct form, as detailed below. 

\subsection{Incoherent time evolution}\label{sec:noise_mapping}
The incoherent part of the Lindblad master equation~(\ref{eq:Lindbladian}) is realized by utilizing incoherent errors
that occur during the time-propagation algorithm.
Incoherent errors have been found~\cite{Cattaneo2023} to explain most of the gate infidelity in IBM-Q devices
similar to the ones used in our work (Sec.~\ref{sec:results}). 
Incoherent errors can be modeled as nonunitary operations after the error-free gates,
as indicated by the replacement
\begin{align}\label{eq:noise_insertion}
    \hat U  &\rightarrow  {\cal N} {\cal U} \, .
\end{align}
On the right-hand side, the unitary gate is represented as the superoperator ${\cal U}$
and the effect of the incoherent error is described by the Kraus operator ${\cal N}$.
It is conveniently expressed in the form
\begin{align}\label{eq:noise_as_lindbladian}
{\cal N} = e^{t_{G}\mathcal{L}_{N}} \, ,
\end{align}
where $\mathcal{L}_{N}$ is a Lindbladian (containing only the incoherent part) and $t_G$ is the physical gate time.

We now map incoherent gate errors to an effective time-independent Lindbladian, or, 
in other words: we determine how the noise behaves in the simulated system.
The effective Lindbladian, in turn, should correspond 
to the incoherent part of Eq.~(\ref{eq:Lindbladian}).
To do this,
the sequence of noisy gates inside one Trotter step (the term inside the square brackets of Eq.~(\ref{eq:time_propagation})) is brought into the form
\begin{align}\label{eq:noise_mapping_derivation}
\Pi_k\Pi_l \left[{\cal N}_{kl} {\cal U}_{kl}\right] &= \Pi_k\Pi_l \left[e^{t_{G,kl} \mathcal{L}_{N_{kl}}}{\cal U}_{kl}\right]\approx e^{\tau \mathcal{L}_\textrm{eff}} \, ,
\end{align}
where
to the lowest order in~$\tau$ 
we have~\cite{Fratus2022}
\begin{align}\label{eq:noise_mapping_result}
\mathcal{L}_{\text{eff}} = \mathcal{L}_{H} + \sum_{kl} \frac{t_{G,kl}}{\tau} \bar{\mathcal{L}}_{N_{kl}} \, .
\end{align}
Here, the first term on the right-hand side
corresponds to the coherent time evolution (commutator with $\hat H$),
whereas the second term has all the noise collected during the computation.
The Lindbladian~$\mathcal{L}_{\text{eff}}$ describes how the noise behaves in the simulated system.

To obtain Eq.~(\ref{eq:noise_mapping_result}),
one has to commute all noise terms past all the coherent gates appearing after them~\cite{Fratus2022,Leppakangas2023}.
A simple illustrative example of commuting noise through unitary gates is a circuit with two gates~$\hat U_2 \hat U_1$ and noise,
\begin{align}
    {\cal N}_2 {\cal U}_2 {\cal N}_1 {\cal U}_1  &= {\cal N}_2 {\cal N}_1' {\cal U}_2 {\cal U}_1 \, ,
\end{align}
where the transformed noise is
\begin{align}
    {\cal N}'_1 &= {\cal U}_2 {\cal N}_1 {\cal U}_2^{-1} \, .
\end{align}
This unitary transformation of a noise superoperator implies, in practice, a unitary transformation of the ``normal'' noise operators
in the Lindbladian of Eq.~(\ref{eq:noise_as_lindbladian}).
An important example of this is the effect of the X-gates, which flip the definition of the qubit versus the spin states.
In turn, this makes (for instance) physical qubit damping
look like spin excitation in the simulated system, since
\begin{align}
\hat\sigma_x \hat\sigma_-\hat\sigma_x &= \hat\sigma_+  \label{eq:damping_to_heating} \, .
\end{align}
(The effect of several other gates is analyzed in Ref.~\cite{Leppakangas2023}.)
Under the assumption of a small noise probability per gate, all such transformed Lindbladian terms can be collected together,
and the final effective Lindbladian will be in the form of Eq.~(\ref{eq:noise_mapping_result}).
Also note the rescaling of the noise operators by factors $t_{G,kl}/\tau$, i.e.,
by the relation between the physical gate times~$t_{G,kl}$ and the simulated time-step~$\tau$.
Important is that one is allowed to neglect transformations due to commuting small-angle rotations, or complete decomposition blocks,
since their effect is higher order in~$\tau$.
An automated package doing this numerically for arbitrary circuits is available in Ref.~\cite{qoqo_files_online}.

\begin{figure}
\begin{center}
\includegraphics[width=\columnwidth]{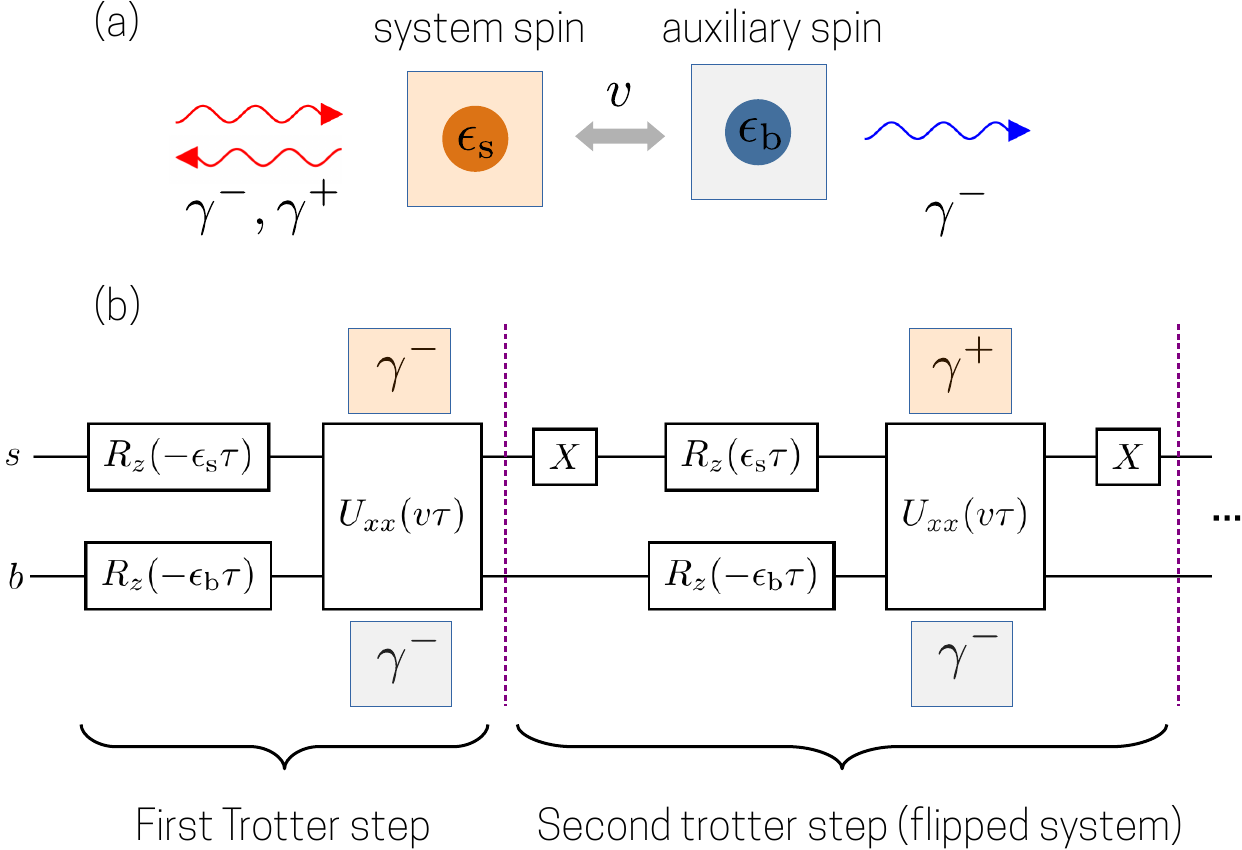}
\end{center}
\caption{
Example of an open quantum system and a corresponding quantum algorithm.
(a) A system spin couples to one auxiliary spin, with the coherent time evolution described by the Hamiltonian
$\hat H = -\frac{\epsilon_\textrm{s}}{2}\hat \sigma_z^{\textrm{s}} - \frac{\epsilon_\textrm{b}}{2}\hat \sigma_z^{\textrm{b}} + v\hat \sigma_x^{\textrm{s}}\hat \sigma_x^{\textrm{b}}$.
Additionally, spin damping and excitation (heating) acts on the system spin, whereas only damping acts on the auxiliary spin.
(b) The corresponding quantum algorithm on a quantum computer with two-qubit gate~$U_{xx}$, with damping acting on both qubits (gate noise).
The X-gates surrounding every second Trotter step are used to  
effectively transform the system-qubit damping to system-spin heating (damping and excitation).
}\label{fig:circuit_ideal}
\end{figure}

\subsection{Tailoring the final algorithm}\label{sec:tailoring_system_bath_physics}
Whereas the coherent part of the Lindblad master equation~(\ref{eq:Lindbladian}) is implemented by standard Trotterization,
the main difficulty lies in finding gates and gate decompositions that simultaneously reproduce the correct incoherent part also.

Our target simulation has an incoherent part, which takes the form of infinite-temperature background for the system, i.e., Pauli noise acting on the system spins, Eq.~(\ref{eq:Lindbladian3}).
There are two central motivations for this goal:
(i) in this form, the system noise drives the system to a fully mixed state and allows for
a physical interpretation of it as a hot background (or equivalently as a constant in the spectral function, see~Appendix~\ref{sec:spectrum_sketch}),
(ii) the effect of the auxiliary spins is now easy to detect. At infinite temperature, all Pauli operators or Pauli products we measure would be zero.
However, since the auxiliary spins lead to cooling of the system, certain Pauli operators or Pauli products will be non-zero.
Additionally, the algorithm we use simultaneously mitigates coherent gate-errors in the system, since it is similar to 
a spin-echo protocol.

To achieve our goal for hardware characterized by qubit damping,
we perform X-flips ($\pi$-rotations) to every second Trotter circuit,
so that system-qubit damping looks like an equal contribution of spin damping and excitation in the simulated system.
In practice, this means that X-gates of system qubits are inserted at the beginning and the end of every second Trotter step.
The flipped circuits have negated angles of system-qubit Z-rotations, as shown in the example in Fig.~\ref{fig:circuit_ideal}(b).
This approach is adequate for the considered form of the Hamiltonian and used hardware, but could
be trivially generalized to symmetrization of other noise types,
as listed in Table~\ref{tab:effect_of_symmetrization}.
It should be noted that
such an algorithm has similarities with randomized compiling~\cite{Wallman2016, Hashim2021, Perrin2024},
which is an alternative way to reach the same goal. 
\begin{table}[ht]
\begin{tabular}{cccc}
\hline
Original noise & X-flips  & Y-flips  & Z-flips  \\
\hline
$\hat\sigma^{-}\equiv Z^{-}$ & X and Y noise & X, Y & no effect \\
$Y^{-}$                & X, Z & no effect & X, Z \\
$X^{-}$                & no effect & Y, Z & Y, Z \\
\hline
\end{tabular}
\centering
\caption{
The effect of the symmetrization algorithm, with different possible spin-flip directions, to the form of the simulated noise.
If the flip is performed around the axis that is different from the direction of the lowering operator, the original noise transforms into Pauli noise with two components.
}
\label{tab:effect_of_symmetrization}
\end{table}

\begin{figure}
\begin{center}
\includegraphics[width=\columnwidth]{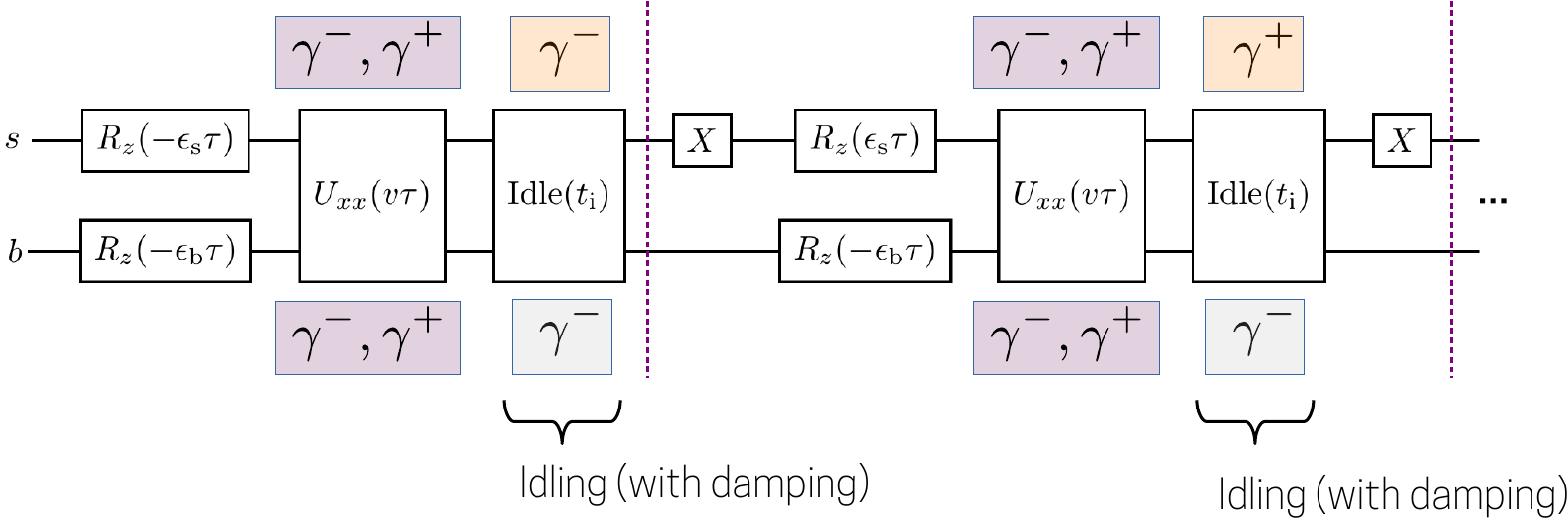}
\end{center}
\caption{
The Trotter circuit of the same global system as in Fig.~\ref{fig:circuit_ideal},
but the contribution of damping to the effective noise model is increased by additional idling at the end of the Trotter steps.
This is needed if the gate noise is depolarization (or Pauli noise),
which cannot be utilized for simulating system cooling.
This type of algorithm was used in IBM-Q demonstrations involving in total two and four qubits respectively (Secs.~\ref{sec:1system1bath} and~\ref{sec:2system2bath}).
}\label{fig:circuit_ibm}
\end{figure}

Unlike the decoherence of the system spin, the auxiliary-spin noise must be predominantly damping,
so that it drives the auxiliary spins towards their ground states, enabling the simulated cooling.
A central requirement to be satisfied here
is that we need to use hardware and gates that are characterized by qubit damping.
According to our noise analysis~(Appendix~\ref{sec:NoiseAnalysis}), 
the inherent noise of IBM-Q devices is characterized by qubit damping and dephasing, as expected for superconducting quantum computers.
However, we find that the application of two-qubit gates (CNOTs)
is accompanied by fully randomizing noise to the control and target qubits.
This result, especially for the CNOT control qubit, is, at first sight, surprising, but
can be explained by a closer look at the implementation of the CNOT gates,
which include an echoed cross-resonance protocol~\cite{Sheldon_2016},
similar to the X-flip protocol described above, effectively leading to symmetrized noise.
Nevertheless, we find that
the quantum algorithm can work also on IBM quantum computers,
since damping during (unavoidable) qubit idling can be utilized.
Qubit idling appears, for instance, when other qubits are acted on or when calling the identity gate, i.e., qubit sleep.
In practice, this means that, for large global systems with unavoidable significant idling time
(due to the form of the considered Hamiltonian),
the effective Lindbladian ${\cal L}_\textrm{ eff }$ is of the desired form.
For small global systems, however, we alter the algorithm by placing an additional sleep phase in the Trotter circuit,
as visualized in Fig.~\ref{fig:circuit_ibm}. This leads to the successful simulations presented in the next section.

\section{Results: System-bath on IBM-Q}
\label{sec:results}

\begin{figure*}
\begin{center}
\includegraphics[width=2\columnwidth]{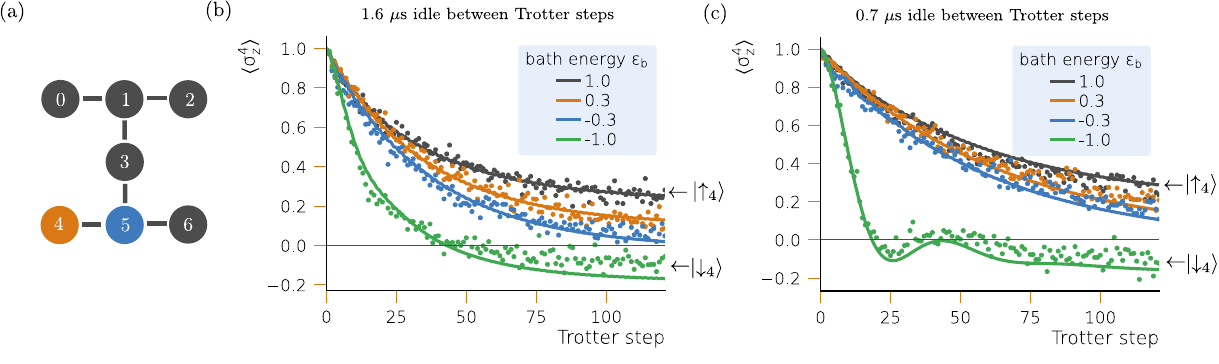}
\end{center}
\caption{
Time-evolving a system-bath model consisting of one system spin coupled to one auxiliary spin on the IBM-Q Nairobi device.
(a) The selection of the system (qubit~4) and bath~(qubit~5) on the IBM-Q Nairobi device, representing the system and the auxiliary spin.
(b) The measured (dots) and numerically simulated (solid lines) expectation values of system $\hat \sigma^\textrm{s}_z$
when starting from the state $\left| \uparrow_\textrm{s},\uparrow_\textrm{b}\right\rangle$
for four different auxiliary-spin energies~$\epsilon_\textrm{b}$,
while fixing the system energy to $\epsilon_{s}=1.0$.
The system steady-state prefers populating the ground state (expectation value above zero) or the excited state
(expectation value below zero), depending on the sign of~$\epsilon_\textrm{b}$.
The measured data corresponds to an average over~$1024$~measurements,
whereas the numerical simulation corresponds an average over infinite measurements.
We use idling phases of~$1.6~\mu$s in each Trotter circuit to enhance qubit damping~(Fig.~\ref{fig:circuit_ibm}).
We use a system-bath coupling of $v=0.1$ and a Trotter time-step of $\tau = 1.0$.
(c) The same simulation as in~(b), but now performed with idling phases of~$0.7~\mu$s,
leading to a reduced simulated noise and a slower relaxation towards the steady-state.
The numerical simulations include fitting variables that are needed to account for coherent gate errors (Appendix~\ref{sec:appNumericalModel}). 
}\label{fig:1system_1bath}
\end{figure*}

Here we present and analyze our results on running the system-bath quantum algorithm on the IBM-Q devices.
In Sec.~\ref{sec:1system1bath}, we start by confirming in the simplest setup
that system-bath physics can be simulated using inherent noise as proposed.
There we consider a global system consisting of one system spin and one auxiliary spin.
In particular,
we show that the steady-state can be made to prefer the ground state of the system or,
if negating the energy of the bath, we can cause population inversion in the system.
After learning the details of how to optimize circuits for the IBM-Q devices,
we go on to test the approach for larger global systems.
The runs for two system spins and two auxiliary spins are analyzed in Sec.~\ref{sec:2system2bath} and
the runs for three system spins and four auxiliary spins in Sec.~\ref{sec:3system4bath}.
In these simulations, we demonstrate that
the steady-state of the system can be made to prefer ferromagnetic or antiferromagnetic spin ordering,
depending on how the system Hamiltonian is defined.
The circuits submitted to IBM-Q and the details of our numerical simulations
are provided in Appendices~\ref{sec:appCircuits} and~\ref{sec:appNumericalModel}.
They are also available online at Ref.~\cite{qoqo_files_online}.

\subsection{One system and one bath qubit: simulated cooling and population inversion}\label{sec:1system1bath}
We start by confirming that system-bath physics can be simulated using inherent noise as proposed.
We study the case of a single system and a single auxiliary spin,
with coherent time evolution given by
\begin{align}
\hat H &= -\frac{\epsilon_\textrm{s}}{2}\hat\sigma_z^{\textrm{s}} - \frac{\epsilon_\textrm{b}}{2}\hat\sigma_z^{\textrm{b}} + v\hat\sigma_x^{\textrm{s}}\hat\sigma_x^{\textrm{b}} \, ,
\end{align}
where we have set $\hbar=1$ and, for simplicity, have also set the simulated time-step $\tau=1$.
We fix the system energy-splitting $\epsilon_{s}$ and the system-bath coupling $v$
but sweep over the bath energy-splitting~$\epsilon_{b}$, i.e.,
the position of the peak in the effective bath spectral density (see Fig.~\ref{fig:spectral_function} in the appendix).
We then expect that the steady-state of the system can be changed by the energy of the auxiliary spin:
if $\epsilon_\textrm{s}=\epsilon_\textrm{b}>0$, the bath should absorb energy from the system,
and if $\epsilon_\textrm{b}\rightarrow-\epsilon_\textrm{b}$,
the system excited state should be preferred.

These results are confirmed in the runs shown in Fig.~\ref{fig:1system_1bath} performed on the IBM-Q Nairobi device.
The selection of system and bath qubits, representing the system and auxiliary spins, is visualized in Fig.~\ref{fig:1system_1bath}(a).
In Figs.~\ref{fig:1system_1bath}(b-c),
we study the relaxation of the expectation value $\langle \hat\sigma^\textrm{s}_z\rangle$ towards the steady-state
for four different bath energies~$\epsilon_{b}$.
We start the simulation always from the state
$\lvert \uparrow_\textrm{s},\uparrow_\textrm{b} \rangle \equiv \lvert \uparrow_4,\uparrow_5 \rangle$,
run the algorithm for a given number of Trotter steps~$m$ (simulated time $t=m\tau$),
measure  $\hat\sigma^\textrm{s}_z$, and average over $N_\textrm{m}= 1024$~repetitions (for each~$m$).
As discussed in Sec.~\ref{sec:tailoring_system_bath_physics},
we add an idling phase to each Trotter circuit to increase the spin damping:
in Fig.~\ref{fig:1system_1bath}(b) the idling phase lasts $1.6~\mu$s
and in Fig.~\ref{fig:1system_1bath}(c)~$0.7~\mu$s.
We confirm that the system steady-state prefers to populate the ground state or the excited state,
depending on the sign of the implemented auxiliary-spin energy.
The behavior is reproduced well by numerical simulations of the quantum computer (solid lines).
The numerical simulations include fitting variables (see Appendix~\ref{sec:appNumericalModel}) 
which are needed to adjust to coherent gate errors and possibly changed T1 and T2 times of the qubits after the calibration (Appendix~\ref{sec:NoiseAnalysis}). 

We find that the idle time between~$1~\mu$s and $2~\mu$s in each Trotter step was the ``sweet spot'' in multiple simulations:
using longer or shorter idle phases made the separation between the curves smaller.
The total idle times were larger or comparable to $T_1$ decay times of the qubit
($\sim 400~\mu$s and~$\sim 200~\mu$s compared to~$\sim 100~\mu$s).
It should, however, be noted that
especially in the simulation of~Fig.~\ref{fig:1system_1bath}(c), the system has yet not fully reached the steady-state,
which was due to the limitation of the overall circuit length set by the IBM-Q devices.

The final state of the system is a mixed state. 
To provide a practical measure of mixing, we calculate the purity of the system, $\textrm{Tr}[\hat \rho_\textrm{s}^2]$, 
which is obtained from the numerical simulations. A pure state gives the upper limit~$1$, whereas the fully mixed state gives the 
lower bound~$1/d$, where~$d$ is the dimension of the Hilbert space. 
For the idling time of $1.6~\mu$s and the cases $\epsilon_\textrm{b}=1$ and $\epsilon_\textrm{b}=-1$,
we obtain the purities $0.5312$ and $0.5139$, respectively, 
indicating rather strong mixing. 

In order to improve the efficiency of the cooling, 
one needs to reduce the qubit noise that heats the system. 
We find that an important limiting factor of cooling 
here is the depolarizing noise of the two-qubit gates: ideally the noise after two-qubit gates would be just qubit damping.
We also find that the strong dephasing during idling is not beneficial: ideally the qubit coherence would be $T_1$-limited.
However, the main limiting factor is the presence of (any form of) system-qubit noise,
whose effect is always significant if noise characteristics of the system and bath qubits are identical.
This means that for efficient cooling the system spins should be represented by low-noise qubits.
Only in this case, we can expect that the maximal steady-state variation between~$\pm 1$ (and purities close to~$1$)
can be approached.

\subsection{Two system and two bath qubits: ferromagnetic and antiferromagnetic steady-states}\label{sec:2system2bath}
\begin{figure*}[t]
\begin{center}
\includegraphics[width=2\columnwidth]{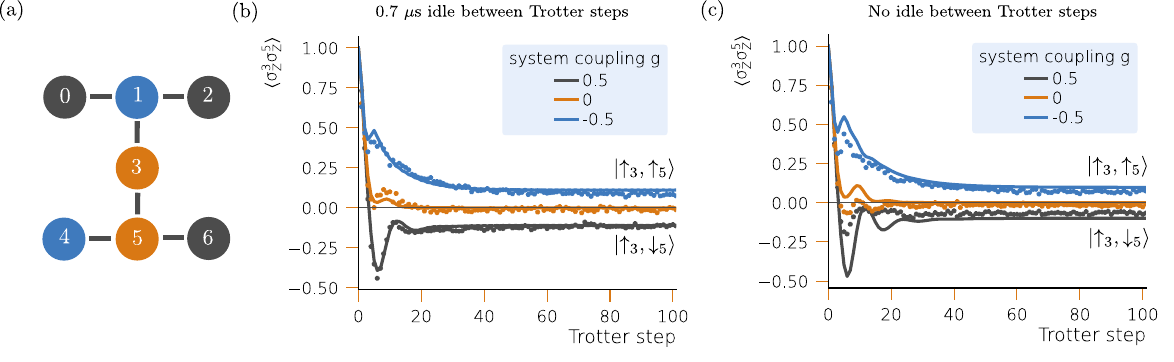}
\end{center}
\caption{
Time-evolving a system-bath model consisting of two system spins coupled to two auxiliary spins on the IBM-Q Lagos device.
(a) The selection of the system qubits and bath qubits on the IBM-Q Lagos device.
(b) The measured (dots) and numerically simulated (solid lines) expectation values of
the system operator $\hat \sigma_z^{\textrm{s},1}\hat\sigma_z^{\textrm{s},2}\propto \hat H_\textrm{S}$,
revealing the preferred spin ordering in the steady-state.
We always start the simulation from all spins being in the state $ \left| \uparrow \right\rangle$
and vary the inner-system spin-spin coupling~$g$.
The interaction with the auxiliary-spin bath makes the system prefer lowest-energy system eigenstates in the steady-state, which are defined by the sign of the coupling~$g$.
When we set the coupling $g=0$, the measured steady-state value is close to zero.
In this simulation we use two (additional) idling phases of $0.7~\mu$s in each Trotter circuit, similarly to Fig.~\ref{fig:circuit_ibm}.
(c) The same simulation as~(b), but here performed without the additional idling phases.
Here we also find a significant separation of the steady-states, since unavoidable idling of qubits already exists during the implementation of spin-interactions in the quantum algorithm.
In both simulations, the measured data corresponds to an average over~$\approx 16000$~measurements and
the numerical simulation to an average over infinite measurements,
and we have system-bath couplings $v=0.2$, bath energies $\epsilon_\textrm{b}=1.0$, and a Trotter time-step of $\tau= 1.0$.
}\label{fig:2system_2bath}
\end{figure*}

In the next simulation, we increase the size of the global system and consider two system spins coupled to two auxiliary spins.
We include two auxiliary spins instead of one to increase the cooling efficiency, although here the theoretical benefit is not significant. 
To address certain limiting factors, we ideally select bath qubits with longer $T_1$-times compared to those of the system qubits.
The conditions for this selection were not available at the Nairobi IBM-Q device, but they were available at IBM-Q Lagos.
The selection of system and bath qubits on the IBM-Q Lagos device is shown in Fig.~\ref{fig:2system_2bath}(a).
The coherent time evolution is described by the Hamiltonian
\begin{align}
\hat H &=  g\hat\sigma_z^{\textrm{s},1}\hat\sigma_z^{\textrm{s},2} -\frac{\epsilon_\textrm{b}}{2}\hat\sigma_z^{\textrm{b},1} - \frac{\epsilon_\textrm{b}}{2}\hat\sigma_z^{\textrm{b},2} + v\hat\sigma_x^{\textrm{s},1}\hat\sigma_x^{\textrm{b},1}  + v\hat\sigma_x^{\textrm{s},2}\hat\sigma_x^{\textrm{b},2} \, .
\end{align}
The coupling between the system spins is longitudinal, the system-bath coupling is transverse, and the system-spin splittings are set to zero.
The eigenstates of the system Hamiltonian $\hat H_\textrm{S}=g\hat\sigma_z^{\textrm{s},1}\hat\sigma_z^{\textrm{s},2}$
are the ferromagnetic orderings $\lvert \uparrow_{\textrm{s},1},\uparrow_{\textrm{s},2}\rangle  \equiv \lvert \uparrow_3,\uparrow_5 \rangle$ and
$\lvert \downarrow_{\textrm{s},1},\downarrow_{\textrm{s},2}\rangle$
with the eigenenergy~$g$,
and the antiferromagnetic orderings
$\lvert \uparrow_{\textrm{s},1},\downarrow_{\textrm{s},2}\rangle$ and $\lvert \downarrow_{\textrm{s},1},\uparrow_{\textrm{s},2}\rangle$
with the eigenenergy~$-g$.
The energy difference between these two orderings is $2g$.
To optimize the energy-absorption efficiency by the bath,
we set both auxiliary-spin energies to $\epsilon_\textrm{b}=2\vert g \vert$.
We then expect that, depending on the sign of the implemented $g$,
one of the two orderings to be preferred in the steady-state, i.e., ferromagnetic for $g<0$ and antiferromagnetic for $g>0$.

In Fig.~\ref{fig:2system_2bath}(b-c), we show that the expected behavior is obtained on the IBM-Q device.
We plot the result for the measured system operator $\hat\sigma^\textrm{s,1}_z\hat\sigma^\textrm{s,2}_z$
during simulated relaxation to the steady-state.
The measured average value of $\langle\hat\sigma^\textrm{s,1}_z\hat\sigma^\textrm{s,2}_z\rangle$ reveals which spin ordering is preferred.
We start the simulation always from 
all spins being in the state $\lvert \uparrow \rangle$ and run the algorithm
until the steady-state is reached.
In Fig.~\ref{fig:2system_2bath}(b) we use $0.7~\mu$s sleep phase in each Trotter circuit
and in Fig.~\ref{fig:2system_2bath}(c) we do not use any sleep phase.
We perform $N_\textrm{m}\approx 16000$ measurements at each time-step, reducing the measurement uncertainty considerably~($\sim 1/\sqrt{N_\textrm{m}}$).
We confirm the expected shift of the average of $\langle\hat\sigma^\textrm{s,1}_z\hat\sigma^\textrm{s,2}_z\rangle$
to the direction which reduces the system energy:
for $g<0$ the ferromagnetic ordering is preferred and for $g>0$ the antiferromagnetic ordering is preferred.
On the other hand, when we set the coupling to $g=0$, the measured expectation value lies in between these results, close to zero.
The purities of the final states are according to numerical simulations $0.2530$ ($g=\pm 0.5$), $0.25$ ($g=0$) for the case of $0.7~\mu$s sleep phases
and $0.2524$ ($g=\pm 0.5$), $0.25$ ($g=0$) for no sleep phases.
Notable is that
we find a significant separation of the steady-state values also without the additional idling phases, see Fig.~\ref{fig:2system_2bath}(c).
This is due to the unavoidable idling of qubits during execution of the quantum algorithm.
This occurs during execution of CNOT gates, with gate-times varying between $0.1~\mu$s and $0.5~\mu$s.
We also observe that the separation of the curves in the steady-state is usually largest for (additional) idling phases between~$0.5~\mu$s and~$1.0~\mu$s.

\subsection{Three system and four bath qubits: stabilized long-range correlations}\label{sec:3system4bath}
\begin{figure*}
\begin{center}
\includegraphics[width=1.8\columnwidth]{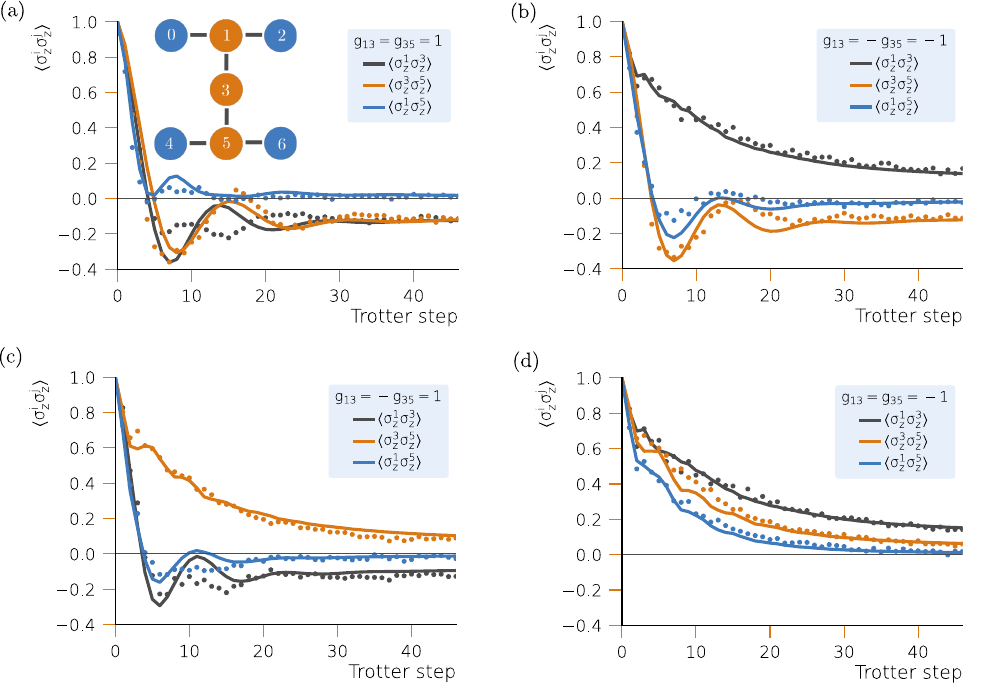}
\end{center}
\caption{
Time-evolving a system-bath model consisting of three system spins coupled to four auxiliary spins on the IBM-Q Nairobi device.
(a) The insert shows the selection of the system qubits and bath qubits on the IBM-Q Nairobi device.
(a-d) The measured (dots) and numerically simulated (solid lines) expectation values of
the system correlations $\langle \hat\sigma^{\textrm{s},i}_z \hat\sigma^{\textrm{s},i'}_z\rangle$ during relaxation to a steady-state
for all (four) different sign-combinations of inner-system couplings $g_{12}$ and $g_{23}$.
We always start the simulation from the state where all spins are in the state $\left| \uparrow \right\rangle$.
The auxiliary spins absorb energy from the system and the steady-state prefers reduced system energy,
which leads to finite steady-state values of the spin-spin correlations.
The measured data corresponds to an average over~$\approx 16000$~measurements, whereas
the numerical simulation corresponds to an average over infinite measurements.
We use inner-system couplings $\vert g_{12}\vert = \vert g_{23}\vert = 0.5$, system-bath couplings $v=0.2$, and a Trotter time-step of $\tau= 1.0$.
No sleep phases (additional idling times) are introduced to the Trotter circuits.
The numerical simulations include fitting variables that are needed to account for coherent gate errors (Appendix~\ref{sec:appNumericalModel}). 
}\label{fig:3system_4bath}
\end{figure*}

In the last simulation, we increase the size of the global system to the maximum allowed by the used IBM-Q devices:
we divide the global system into three system spins and four auxiliary spins, as shown in the insert of Fig.~\ref{fig:3system_4bath}(a).
The inclusion of maximal number of auxiliary spins is again to increase the cooling efficiency,
although the theoretical benefit of having more than two auxiliaries, one connected at each end, is not significant.
The coherent time evolution is described by the Hamiltonian
\begin{align}\label{eq:Hamiltonian3system4bath}
\hat H &=  g_{12}\hat\sigma_z^{\textrm{s},1}\hat\sigma_z^{\textrm{s},2} + g_{23}\hat\sigma_z^{\textrm{s},2}\hat\sigma_z^{\textrm{s},3} \nonumber \\
&-\frac{\epsilon_{b}}{2}\left(\hat\sigma_z^{\textrm{b},1} + \hat\sigma_z^{\textrm{b},2} + \hat\sigma_z^{\textrm{b},3} + \hat\sigma_z^{\textrm{b},4}\right) \nonumber \\
 & + v\sigma_x^{\textrm{s},1}\left(\hat\sigma_x^{\textrm{b},1}  + \hat\sigma_x^{\textrm{b},2}\right) + v\hat\sigma_x^{\textrm{s},3}\left(\hat\sigma_x^{\textrm{b},3} + \hat\sigma_x^{\textrm{b},4}\right) \, ,
\end{align}
modeling longitudinal nearest-neighbor coupling in the system and
transverse nearest-neighbor coupling between the system and bath. For simplicity, we have again set the system energies~$\epsilon_\textrm{s}$ to zero.
The eigenstates of the system Hamiltonian [the first line on the right-hand side of Eq.~(\ref{eq:Hamiltonian3system4bath})]
are product states of $\lvert\uparrow_{\textrm{s},i}\rangle$ and $\lvert\downarrow_{\textrm{s},i} \rangle$ of system spins~$i$.
The lowest-energy eigenstate is degenerate and is determined by the signs of the couplings $g_{ii'}$:
it has opposite spin-directions for neighboring spins with positive coupling,
and vice versa for the negative coupling.
The degeneracy is due to the invariance of the energy when flipping all spin-directions simultaneously.
The lowest eigenenergy is always $E_\textrm{s} = -\vert g_{12}\vert  -\vert g_{23}\vert $.

For optimizing the energy absorption by the auxiliary-spin bath,
the system energy-changes under transitions caused by the bath are essential.
Here, the transition is always a flip of system spin~1 or~3 with a corresponding energy change of $\pm 2\vert g_{12}\vert $ or $\pm 2\vert g_{23}\vert$.
The sign is defined by the flip direction.
Since we consider here the case of equal-magnitude system couplings, $\vert g_{12}\vert = \vert g_{23}\vert = g$,
it is then optimal to choose all auxiliary-spin energies to be
$\epsilon_{\textrm{b}} = 2 g$ to maximize the energy absorption efficiency by the bath.

In Fig.~\ref{fig:3system_4bath}(a-d),
we plot the result for the measured $\langle \hat\sigma^{\textrm{s},i}_z\hat\sigma^{\textrm{s},i'}_z \rangle$ expectation value
during the relaxation to the steady-state for all four different sign combinations of the couplings~$g_{12}$ and~$g_{23}$.
Each plot includes three curves, corresponding to the three possible system-index pairs~$(i,i')$.
We again start the simulations from all spins being in the state $\lvert\uparrow\rangle$
and run the algorithm until the steady-state is reached.
We perform $N_\textrm{m}\approx 16000$ measurements at each time-step.
We always observe a steady-state value  with the expected sign, i.e., a sign that reduces the average system energy.
For example, for an antiferromagnetic Hamiltonian, the nearest-neighbor spins prefer pointing to opposite directions,
which also means the next-nearest neighbor spins prefer pointing to the same direction, which is indeed measured in Fig.~\ref{fig:3system_4bath}(a).
Also, changing the sign of one of the nearest-neighbor couplings causes a switch in the corresponding spin ordering,
also resulting in that the next-nearest-neighbor spins now prefer pointing to the opposite directions, as also measured in Fig.~\ref{fig:3system_4bath}(b-c)
The measured effect is strongest between nearest-neighbor spins,
but it is also observed between next nearest-neighbor spins, confirmed by the fact that
the measured mean values were always larger than the standard deviation when averaging over the last 15 Trotter steps.
The same result is reproduced also by the numerical modeling of the simulations and can be understood by simple reasoning:
if the probability for coherence between system spins~1~and~2 is~$p$, which is well below~1,
and the probability for coherence between system spins~2~and~3 is also~$p$,
then the indirect probability for the coherence between spins~1~and~3 can be expected to be $p^2 \ll 1$.
Note that we did not need to introduce any additional sleep phases to the Trotter circuits (to increase simulated bath damping),
since significant qubit idling is inevitable already without them, due to the presence of multiple non-commuting spin-spin interactions in the Hamiltonian.
The average purity for the four different final states was $0.1287$~($\pm 0.001$).

In this specific realization of the open-system model,
the steady-state correlations between the distant system spins remained weak.
In another application, we may want to realize a regime where the long-range coherence is strong,
which means achieving essentially more efficient algorithmic cooling.
In the presented realization,
one specific limiting factor for long-range coherence was the absence of the auxiliary spins connected to the center system-spin, see Fig.~\ref{fig:3system_4bath}(a).
According to numerical simulations, an efficient realization would use auxiliary spins connected to the center spin
that have doubled spin energies, in order to trigger system transitions with energy-level differences~$4g$.
This suggests that, in general, the most optimal configuration involves multiple bath qubits connected to each system qubit, with a bath spectral density that is optimally tailored~\cite{Leppakangas2023}.

The main limiting factor for the long-range coherence is the presence of system-qubit noise (of any form).
This maps to local heating at each system spin, with a total rate that then increases with system size.
In such situation,
the number of auxiliaries needs to be increased with the system size,
in order to compensate for the increasing total decoherence rate~\cite{Mi2023} and maintain (at least) local coherence.
On the other hand, for a very weak local noise, even a single auxiliary spin can be enough to cool the system,
as discussed in Ref.~\cite{Raghunandan_2020} in the case of transverse-field Ising model.
(This result is however model dependent: for the system considered here, multiple auxiliary spins are always needed.)
We can then conclude that for establishing strong long-range correlations, the system qubits need to have higher quality
and the number of available bath qubits needs to increase with the system size.


\section{Conclusion and discussion}
\label{sec:conclusion}
We have demonstrated the utilization of inherent qubit noise in digital quantum simulation
by mapping its effect to physical processes. 
In particular, we have shown that qubit damping
can be exploited to create the key effect of a finite-temperature bath,
namely nonunitary time evolution with its direction defined by energy changes.
This digital quantum simulation creates steady-states that prefer to populate low-energy eigenstates
of the simulated system.
The created steady-state is characterized by non-local qubit-qubit correlations,
which are robust against the noise, since their creation is based on utilizing it.
The state can be maintained, in principle, as long as the algorithm is being executed.
The exact form of the correlations can be engineered by the definition of the Hamiltonian
and the details of the gate decompositions and the time-propagation algorithm.
Our work presents, to the best of our knowledge,
a first demonstration of finite-temperature system-bath physics (and algorithmic cooling) on a publicly available quantum computer.
The circuits we have developed and run on the IBM-Q devices are available online at Ref.~\cite{qoqo_files_online},
together with the shown numerical simulations of noisy quantum computation.

The final goal of this work is to be able to perform either useful simulation of materials and molecules 
by solving relevant system-bath models~\cite{Wouters2016, Sun_2020, Rozenberg1996, Spin_Boson_Rev},
or to map open quantum systems to relevant quantum machine learning algorithms. 
A tool to translate arbitrary system-bath models to noise-utilizing quantum simulations
is ``bath mapping'', described in Ref.~\cite{Leppakangas2023}.
The final quantum algorithm in the bath mapping is analogous to the one presented in this paper,
with the additional level of Hamiltonian-parameter determination so that
the target spectral function is simulated as well as possible.

Finally, long-range multi-party correlations and entanglement, in particular, pose major difficulties for classical simulation methods.
Therefore, for useful applications of system-bath models, devices will need a large number of qubits.
We concluded that, for achieving stronger long-range steady-state correlations,
certain criteria for hardware properties should be reached.
We found that qubit architectures which are at least bilinear or trilinear, $T_1$ limited qubits, and low-noise system qubits would be necessary.
Ideally, variable-angle two-qubit gates would be available on the hardware,
such as the variable iSWAP gates on superconducting quantum computers~\cite{Mi2023}, as they would be particularly useful~\cite{Leppakangas2023},
since they can be expected to be characterized by damping noise instead of fully depolarizing noise.
It can also be expected that the use of control-Z gate decompositions
with tunable qubits and couplers~\cite{Mi2023} would perform better than the CNOT gate decompositions with fixed-frequency qubits available in this work.
The main obstacle for achieving low-temperature quantum simulations
is, however, the presence of noise in the system qubits, which, in the used algorithm, mapped to the presence of an infinite-temperature background environment.
Therefore, achieving long-range multi-party correlations and entanglement would be most efficiently reached by 
having system qubits (not bath qubits) of substantially better quality. 


\acknowledgments{
This work was supported by the German Federal Ministry of Education and Research, through projects Q-Exa (13N16065) and QSolid (13N16155).
}

\bibliography{spinboson}


\appendix


\section{Description using bath spectral function}\label{sec:spectrum_sketch}
\begin{figure}
\begin{center}
\includegraphics[width=\columnwidth]{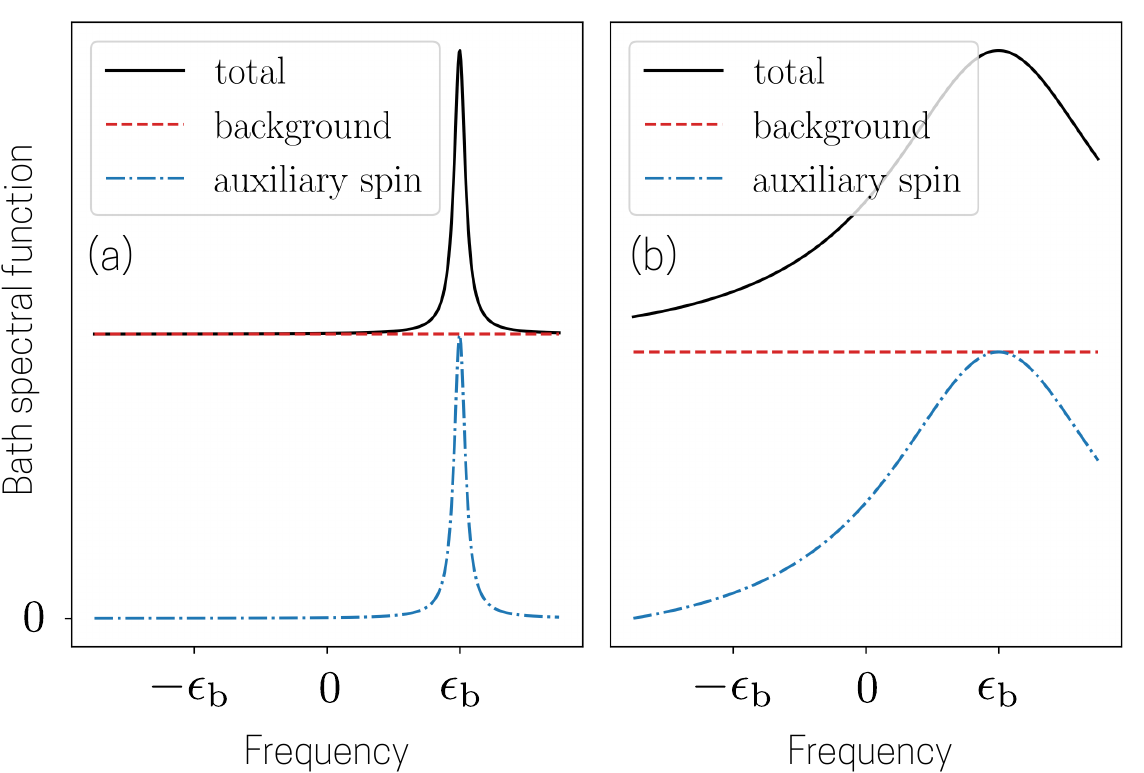}
\end{center}
\caption{
The form of the bath spectral function (solid lines).
We consider the case of one auxiliary spin with damping (a) $\gamma_- = 0.1\epsilon_\textrm{b}$, 
(b) $\gamma_- = 2\epsilon_\textrm{b}$, 
system-bath coupling $v=\gamma_-/2$,
and the use of noise-symmetrization algorithm (see Sec.~\ref{sec:implementation_on_hardware})
on a hardware with homogeneous qubit damping rates.
The spectral function describes how the environment
of the system can absorb (positive energies) and emit (negative energies) energy.
}\label{fig:spectral_function}
\end{figure}

We will now roughly sketch how the cooling and heating effects can interplay and what are the fundamental limitations of the cooling efficiency in this setup.
For this we write down a spectral function for the bath, describing transitions between energy-levels of the system induced by the auxiliary-spin bath and the background.
The bath spectral function describes how the bath absorbs (positive energies) and emits (negative negative) energy as a function of the energy changes of the system~$\omega$.
(Note that we have put $\hbar=1$ and thereby work in units where energies are equal to frequencies.)
In the case of one auxiliary spin, the spectral function is of the simple form~\cite{Leppakangas2023}
\begin{align}\label{eq:spectrum_qualitative}
S(\omega) &= \gamma_\textrm{background} + v^2\frac{\gamma_-}{(\gamma_-/2)^2 + (\omega - \epsilon_\textrm{b})^2} \, ,
\end{align}
which  consists of two parts:
the constant $\gamma_\textrm{background}$ (which will be proportional to the system-qubit noise)
and the Lorentzian at the simulated auxiliary-spin energy~$\epsilon_\textrm{b}$, with the subscript \textit{b} indicating \textit{bath},
with broadening defined by the damping rate of the auxiliary spin~$\gamma_-$,
and area $2\pi v^2$ defined by the system-bath coupling~$v$~[see Eq.~(\ref{eq:coupling_Hamiltonian})].
Most important for our purposes is that the spectral function is not symmetric around the zero energy, i.e.,
the total bath can absorb energy more than it emits. 
The form of the bath spectral function is visualized in Fig.~\ref{fig:spectral_function}.

For couplings $v\lesssim \gamma_-$,
we can make a rough estimate for the system steady-state populations
(considering now a single-spin system), which approximately satisfy
\begin{align}\label{eq:population_ration}
\frac{p_1}{p_0} &\approx \frac{S(-\epsilon_\textrm{s})}{S(\epsilon_\textrm{s})} \, ,
\end{align}
where $p_{0(1)}$ is the population of the state $\vert \uparrow\rangle$ ($\vert \downarrow\rangle$) with energy $\epsilon_{\textrm{s}}/2$
($-\epsilon_{\textrm{s}}/2$), with the subscript \textit{s} indicating \textit{system}.
Assuming that $\epsilon_{\textrm{b}}>0$,
the system is then more probable to populate the lower energy eigenstate.
Note that if we instead simulate $\epsilon_{\textrm{b}}<0$, the auxiliary-spin bath drives the system to population inversion.
In Fig.~\ref{fig:spectral_function}, we have plotted the bath spectral function of Eq.~(\ref{eq:spectrum_qualitative})
corresponding to the coupling $v= \gamma_-/2$ and $\gamma_\textrm{background}=\gamma_-$,
the latter corresponding to the use of the symmetrization algorithm
(see Sec.~\ref{sec:implementation_on_hardware}) with homogeneous qubit damping.
Here, the maximum of the bath-qubit peak equals the constant background rate $\gamma_\textrm{background}$.
According to Eq.~(\ref{eq:population_ration}),
we then have $p_1/p_0\approx 1/2$, assuming $\epsilon_\textrm{b}=\epsilon_\textrm{s}\gg \gamma_-$,
which is close to the exact solution.
An analysis for couplings beyond the validity of  Eq.~(\ref{eq:population_ration})
gives that the maximal population difference is $p_1/p_0 \gtrsim 1/3$ 
achieved in the parameter regime $\epsilon_\textrm{b},v\gg \gamma_-$.
The perfect ground-state population ($p_1/p_0=0$)
can be approached in the limit of vanishing system noise ($\gamma_\textrm{background}\rightarrow 0$).

We note that the spectral weight at negative frequencies may be reduced by
spectral-function filtering implemented via time-dependent system-bath couplings, as shown in Ref.~\cite{Lloyd2024}.
It may also be possible to do this effectively via calculating the effect of negative-valued spectral peaks (at wished locations)
by doing simulations for different system-bath couplings and performing an analytic continuation, as discussed in Ref.~\cite{Lambert2024}.

\section{Noise on IBM-Q}
\label{sec:NoiseAnalysis}

The quantum circuits need to be tailored for the available gates and the characteristics of the device noise.
In this Appendix, we present all essential calibration data retrieved from IBM-Q at the time of the simulations
and results of our additional noise tomography. In the additional analysis
we do not aim to full knowledge of the noise characteristics,
which would need performing a full quantum process tomography
(see for example Refs.~\cite{Poyatos_1997, Chuang_1997}),
but probe the most essential additional information for our purposes, i.e.,
the presence of qubit damping when executing different types of gates and the presence of coherent errors.

\subsection{Calibration data}
In Tables~\ref{tab:calibtration_data1} and~\ref{tab:calibtration_data2}
we list all essential calibration data given by IBM-Q, namely
gate errors, qubit $T_1$~decay and $T_2$~dephasing times, and gate times.
This information also defines parameters of our numerical simulations, described more detailed in Sec.~\ref{sec:appNumericalModel}.
Additionally, we note that
the measurement errors of the used IBM-Q devices varied usually between 1-5 percent.
Such small errors do not play a significant role in explaining the observed behavior and were neglected in our analysis.
Finally, the transmon qubits used in IBM-Q devices~\cite{Sheldon_2016} had commonly frequencies close to 5~GHz with anharmonicities of the order -0.3~GHz.
\begin{table}[ht]
\begin{tabular}{c|c|c|c|c|c|c}
            & $X$ or $\sqrt{X}$ error ($10^{-4}$)   & $T_{1}$ ($\mu$s)   & $T_{2}$ ($\mu$s)    \\
\hline
(s) qubit~4 & 4.01   & 101    & 35.9       \\
(b) qubit~5 & 2.69   & 93.0   & 79.8       \\
\hline
(s) qubit~3 & 1.43   & 106    & 130        \\
(s) qubit~5 & 2.79   & 103    & 84.1       \\
(b) qubit~1 & 3.80   & 93.2   & 111        \\
(b) qubit~4 & 4.00   & 101    & 24.9       \\
\hline
(s) qubit~1 & 3.66   & 98.8   & 120        \\
(s) qubit~3 & 3.59   & 126    & 57.0       \\
(s) qubit~5 & 2.78   & 125    & 21.5       \\
(b) qubit~0 & 2.41   & 113    & 31.9       \\
(b) qubit~2 & 3.84   & 54.5   & 110        \\
(b) qubit~4 & 2.53   & 96.5   & 108        \\
(b) qubit~6 & 2.01   & 122    & 220        \\
\hline
\end{tabular}
\centering
\caption{
IBM-Q calibration data of single-qubit gate errors and decoherence times.
The data for the three demonstrations is separated by horizontal lines and was obtained at the time the simulations were run.
Marking (s) means a system qubit and (b) a bath qubit.
All single-qubit gate times were $35.5$~ns, except the (virtual) rotate-Z gate whose gate time is zero.
}
\label{tab:calibtration_data1}
\end{table}

\begin{table}[ht]
\begin{tabular}{c|c|c|c|c|c|c}
                                   & CNOT error ($10^{-2}$) & gate time (ns)  \\
\hline
(sb)    ctrl.~4-trg.~5 & 0.836                    & 363                 \\
\hline
(ss)    ctrl.~3-trg.~5 & 2.26                     & 960                 \\
(sb)    ctrl.~3-trg.~1 & 0.698                    & 334                 \\
(sb)    ctrl.~5-trg.~4 & 1.28                     & 363                 \\
\hline
(ss)    ctrl.~5-trg.~3 & 1.607                    & 277                 \\
(ss)    ctrl.~3-trg.~1 & 0.650                    & 270                 \\
(sb)    ctrl.~1-trg.~0 & 0.927                    & 249                 \\
(sb)    ctrl.~1-trg.~2 & 0.855                    & 427                 \\
(sb)    ctrl.~5-trg.~4 & 0.481                    & 313                 \\
(sb)    ctrl.~5-trg.~6 & 0.608                    & 341                 \\
\hline
\end{tabular}
\centering
\caption{
IBM-Q calibration data of two-qubit gate errors and gate times.
The data for the three demonstrations is separated by horizontal lines and was obtained at the time the simulations were run.
Marking (s) means a system qubit and (b) a bath qubit.
Only data for the gates that were used in the algorithm is shown.
}
\label{tab:calibtration_data2}
\end{table}

\subsection{Probing noise during qubit idling}\label{sec:noise_idling}

\begin{figure*}
\begin{center}
\includegraphics[width=2\columnwidth]{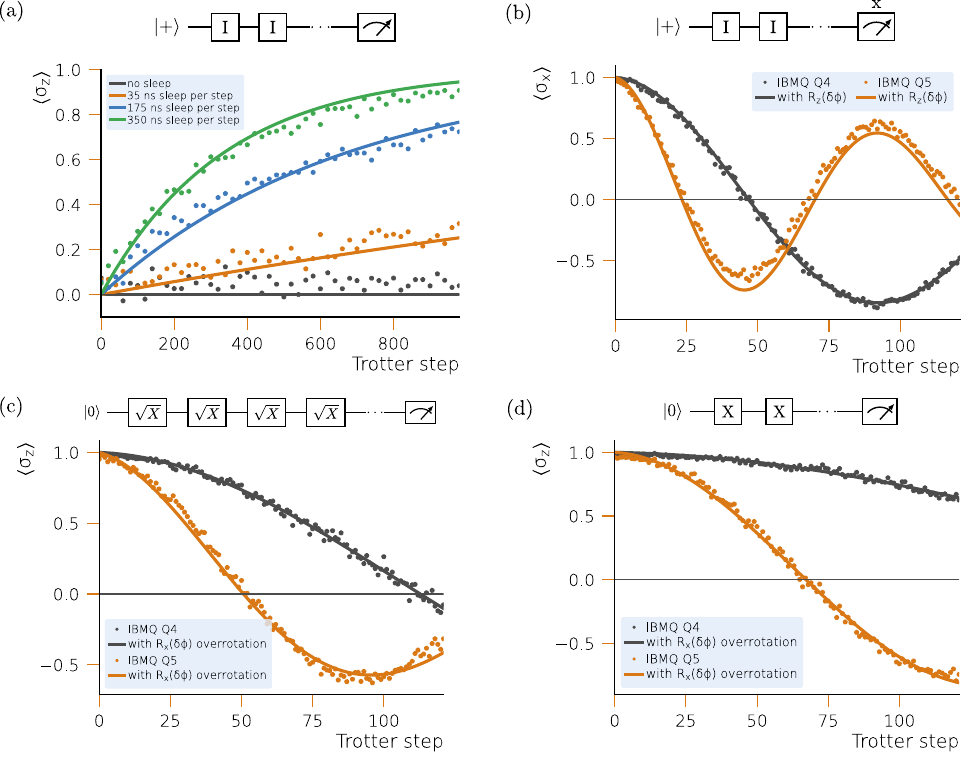}
\end{center}
\caption{
Characterization of single-qubit errors on IBM-Q with circuits whose expectation value differ from 1 only due to errors.
(a)
Probing qubit damping on IBM-Q during qubit sleep (idling).
We initialize the qubit to the eigenstate of $\hat\sigma_x$ and "run" different lengths of sleeping gates per Trotter step
and measure the expectation value of $\hat\sigma_z$ as a function of applied Trotter steps.
The decay rate is consistent with the IBM-Q given $T_1\sim 100~\mu$s (solid lines).
(b)
Probing dephasing of two different qubits by now measuring $\langle\hat\sigma_x\rangle$.
We confirm that idling noise roughly fits to the given $T_2\sim 80~\mu$s.
We also see coherent oscillations, which are most probably due to change of the qubit frequency after the calibration,
leading to an effective $R_z$ gate with angle $\delta \phi=-0.03$ (qubit 4) and $\delta \phi = 0.01 $ (qubit 5) per identity gate (corresponding to $35$~ns qubit-sleep).
(c) Observation of overrotations of the $\sqrt{X}$-gate on IBM-Q.
The oscillations correspond to overrotation of $\delta\phi=-0.01$ (qubit 5) and $\delta\phi=0.003$ (qubit 4) per $\sqrt{X}$-gate.
We also observe weak damping of the oscillations, due to the incoherent error.
(d) Observation of overrotations of $X$-gate on IBM-Q.
}\label{fig:noise_Characterization_damping}
\end{figure*}

\begin{figure*}
\begin{center}
\includegraphics[width=2\columnwidth]{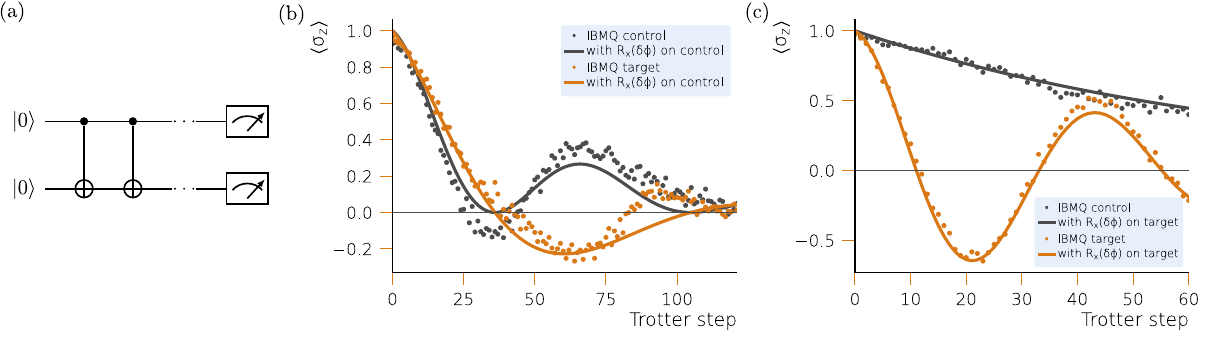}
\end{center}
\caption{
Characterization of two-qubit errors on IBM-Q with circuits whose expectation value differ from 1 only due to errors.
(a) Circuit with 2 CNOT gates. We measure  $\hat\sigma_z$ as a function of applied Trotter steps.
(b) Observation of strong incoherent and coherent error of the CNOT gate for both the control and target qubit on the IBM-Q Nairobi device.
The incoherent noise drives the qubits towards equal population of the qubit states.
The coherent oscillations can be reproduced by additional X- or Y-rotations performed
with angle $\delta\phi=-0.05$ on the control qubit after each CNOT operation. The overrotation on control leads to overroation of the target.
(c) Same run as in (b) but on IBM-Q Quito. We observe an incoherent error on the control and both incohorent and coherent error on the target.
The coherent oscillations can be reproduced by additional X-rotations on the target with angle $\delta \phi = -0.07$, after each CNOT.
}\label{fig:noise_Characterization_CNOT}
\end{figure*}

During idling, the noise of superconducting qubits is expected to be damping and dephasing.
The strengths of these processes can be retrieved from the calibration data ($T_1$ and $T_2$ times),
but can also be probed independently by running corresponding quantum circuits and measurements.
The results may be different, since after the calibration the properties of the quantum computers will change in time.

A probe of the qubit decay rate via state initialization,
use of identity (sleep) gates, and performing qubit measurements is shown in Fig.~\ref{fig:noise_Characterization_damping}(a).
Here we initialize the qubit to the eigenstate of $\hat\sigma_x$ and first monitor the expectation value of $\hat\sigma_z$.
We manifest a decay of the qubit population to its ground state.
The decay rate we obtain from the fitting is consistent with the $T_1$ obtained from the calibration data.
In Fig.~\ref{fig:noise_Characterization_damping}(b), we perform the same protocol but measure $\langle \hat\sigma_x\rangle$.
We observe spurious coherent oscillations,
which can be fitted by a presence of coherent Z-error of angle $\delta \phi=-0.03$ (qubit 4) and $\delta \phi = 0.01 $ (qubit 5) for each identity gate (corresponding to $35$~ns qubit-sleep).
This is most probably due to a frequency shift~$\delta\epsilon$ of the qubit after the calibration~\cite{Klimov_2018,Schl_r_2019},
which effectively introduces a permanent qubit Hamiltonian term $\delta\epsilon\hat\sigma_z/2$.
Such coherent errors change slowly in time but can often be
considered to be constants over one demonstration (as observed here).
The oscillations decay towards zero due to qubit dephasing (finite $T_2$-time),
with significant contribution coming from pure dephasing (instead of qubit damping).

\subsection{Probing the character of single qubit gates}\label{sec:noise_Rx}

The available single qubit gates on IBM-Q are $\sqrt{X}$, $X$, and virtual $R_z$ gates.
A study of the gate $\sqrt{X}$-gate is shown in Fig.~\ref{fig:noise_Characterization_damping}(c).
Here the "Trotter circuit" consists of four $\sqrt{X}$-gates, which generate all together an identity up to a global phase (an operation~$-\hat I$).
In the absence of errors, the expectation value should not differ from the initial value.
However, we observe damped oscillations.
The oscillation period can be fitted by X-overrotations of angle $\delta\phi=-0.01$ (qubit 5) and $\delta\phi=0.003$  per $\sqrt{X}$-gate.
The damping is due to an incoherent error.
The results of the noise characterization of $X$-gates were similar and are shown in Fig.~\ref{fig:noise_Characterization_damping}(d)

The overrotations could be explained with the model of off-resonance rotations~\cite{McKay_2017} (ORR),
which in turn could be due to the influence of higher levels of the qubits.
However, if the resulting overrotation due to ORR over the full cycle of four $\sqrt{X}$ gates is~$\delta\phi$,
then the magnitude of coherent errors in Y- and Z-directions after a single $\sqrt{X}$-gate is~$\sim \sqrt{\delta\phi}$.
This number is consistent with other analyses of the IBM-Q single-qubit error when fitted using ORR errors~\cite{Bultrini2021}.
However, such error would result in a very low fidelity for the single $\sqrt{X}$-gates (of the order~$\delta\phi$)
and is therefore not expected to be the dominating source of the observed error.

The observed oscillations are weakly damped.
The damping is due to an incoherent error after each $\sqrt{X}$-gate, whose magnitude is consistent with inherent decay and dephasing.
The decay is towards zero and thereby the form of the effective noise after four $\sqrt{X}$-gates
is Pauli noise.
The exact form of the noise after one $\sqrt{X}$-gate is not solved in this measurement.
We note that its magnitude is small when compared to the incoherent noise of CNOT gates, but cannot be neglected completely.

\subsection{Characterizing noise of two-qubit gates}\label{sec:noise_CNOT}

The gate decompositions in this work were based on using CNOT gates, available on the IBM-Q devices.
The CNOT gates were the main source of incoherent error in the simulations.
They also came often with significant coherent error.

We have probed the coherent and incoherent errors by Trotter circuits consisting of two CNOT gates, see Fig.~\ref{fig:noise_Characterization_CNOT}.
Again, the expectation value should not deviate from 1 without errors.
For a damping noise after each CNOT, the result should also not change.
On the other hand, in the presence of coherent errors and damping,
the steady-state should still be biased towards the qubit ground state.
We instead observe a fast decay of the qubit populations towards a steady-state where $\langle\hat\sigma_z^i \rangle\approx 0$.
This means that there is roughly an equal amount of qubit damping and excitation, both in the control and in the target.
At first sight, this result is not expected for the control qubit, which should rather stay closer to the ground state.
However, since during the implementation of the CNOT gate the control qubit is operated twice by X-gates~\cite{Sheldon_2016},
the effective noise gets symmetrized to an equal contribution of damping and dephasing,
exactly in the same way as in our noise-symmetrization algorithm [see Eq.~(\ref{eq:damping_to_heating})].

We also observe coherent oscillations (coherent error) on both qubits (control and target) with rather high frequency.
There are many possible origins for this behavior. Firstly,
during the gate~\cite{Sheldon_2016}, both the control and the target are subject to large-angle single-qubit X-rotations,
which can come with overrotations and ORR errors.
Also, the Hamiltonian of the cross-resonance gate has a rich structure of different unwanted terms~\cite{Sheldon_2016},
whose magnitudes will change in time after the calibration.
Also the effect of spurious qubit-frequency shifting may result in similar errors,
due to the long gate times of CNOTs. All together, an exact theoretical modeling of the error would be highly cumbersome,
and therefore we use a simplified coherent error model 
where each CNOT gates comes with additional single-qubit rotations (X or Y).
In Fig.~\ref{fig:noise_Characterization_CNOT}(b), we observe a situation
where it is enough to add rotations only to the control to make also the target oscillate with the observed frequency.
On the other hand, in Fig.~\ref{fig:noise_Characterization_CNOT}(c),
we see an opposite situation, where oscillations exist in the target qubit, caused by coherent error only in the target qubit.
We then observe that also the type of coherent error may vary between different runs.

\subsection{Mapping coherent errors to Hamiltonian disorder}\label{sec:coherent_errors_to_effective_model}
Coherent noise can be mapped to Hamiltonian disorder in the simulated model~\cite{Reiner2018}.
Perhaps the simplest example is
a constant Z-error, due to a spurious qubit-frequency shift [as measured in Fig.~\ref{fig:noise_Characterization_damping}(b)].
This can directly transform to shifts of simulated spin energies
(except when appearing inside decomposition blocks, as discussed below).
We observe this effect in multiple runs of the system-bath algorithm as shifts of the
actually implemented auxiliary spin energies. In particular, long idling times can come with large shifts in the auxiliary spin energies.
Note that X-gates between Trotter steps cancel out coherent Z-errors, but cannot be applied to bath qubits the algorithm to work.

More complicated interplay between coherent errors and gate decompositions
occurs, for example, when coherent errors appear inside CNOT-decompositions.
However, when writing down coherent errors in the discussed (noise-after-gate) form,
it can also here be straightforward to map them to Hamiltonian terms.
For example, an overrotation of control qubit rotation $R_x(g\tau)$ by angle $\delta\phi$
in the implementation of XX interactions (two CNOTs surrounding control-qubit rotation $R_x(g\tau)$)
leads to the shift~$g\rightarrow g+\tilde\phi/\tau$ in the effectively implemented XX-coupling.
Especially the implementation of small Hamiltonian terms (or small Trotter time-steps~$\tau$) is then very fragile against such errors,
since the size of the Hamiltonian disorder scales like~$1/\tau$ relative to the correct Hamiltonian.
Also, target Z-rotations inside the CNOTs (for example due to qubit-frequency jumps, see Sec.~\ref{sec:noise_idling}) induce ZZ-interaction.
It should be noted that also many others types of disorder terms can emerge.
For example, coherent Y-errors of a control qubit between CNOTs leads to effective YX-coupling, and so on.

A table implying which types of Hamiltonian terms do single-qubit coherent errors in different spin directions induce is given in Table~\ref{tab:CliffordTable}.
\begin{table}[ht]
\begin{tabular}{c|ccc}
\hline
Gate, qubit & X  & Y  & Z  \\
\hline
CNOT, 0 & XX & YX & ZI \\
CNOT, 1 & IX & ZY & ZZ \\
CZ, 0   & XZ & YZ & ZI \\
CZ, 1   & ZX & ZY & IZ \\
\hline
\end{tabular}
\centering
\caption{
Coherent errors between CNOT or CZ gates lead to effective Hamiltonian disorder.
The shown (Clifford) table implies which types of Hamiltonian terms emerge from single-qubit coherent errors.
The considered errors can appear either in the control (0) or target (1) qubit and can be rotations around arbitrary spin axes.
}
\label{tab:CliffordTable}
\end{table}


\section{Submitted circuits}
\label{sec:appCircuits}

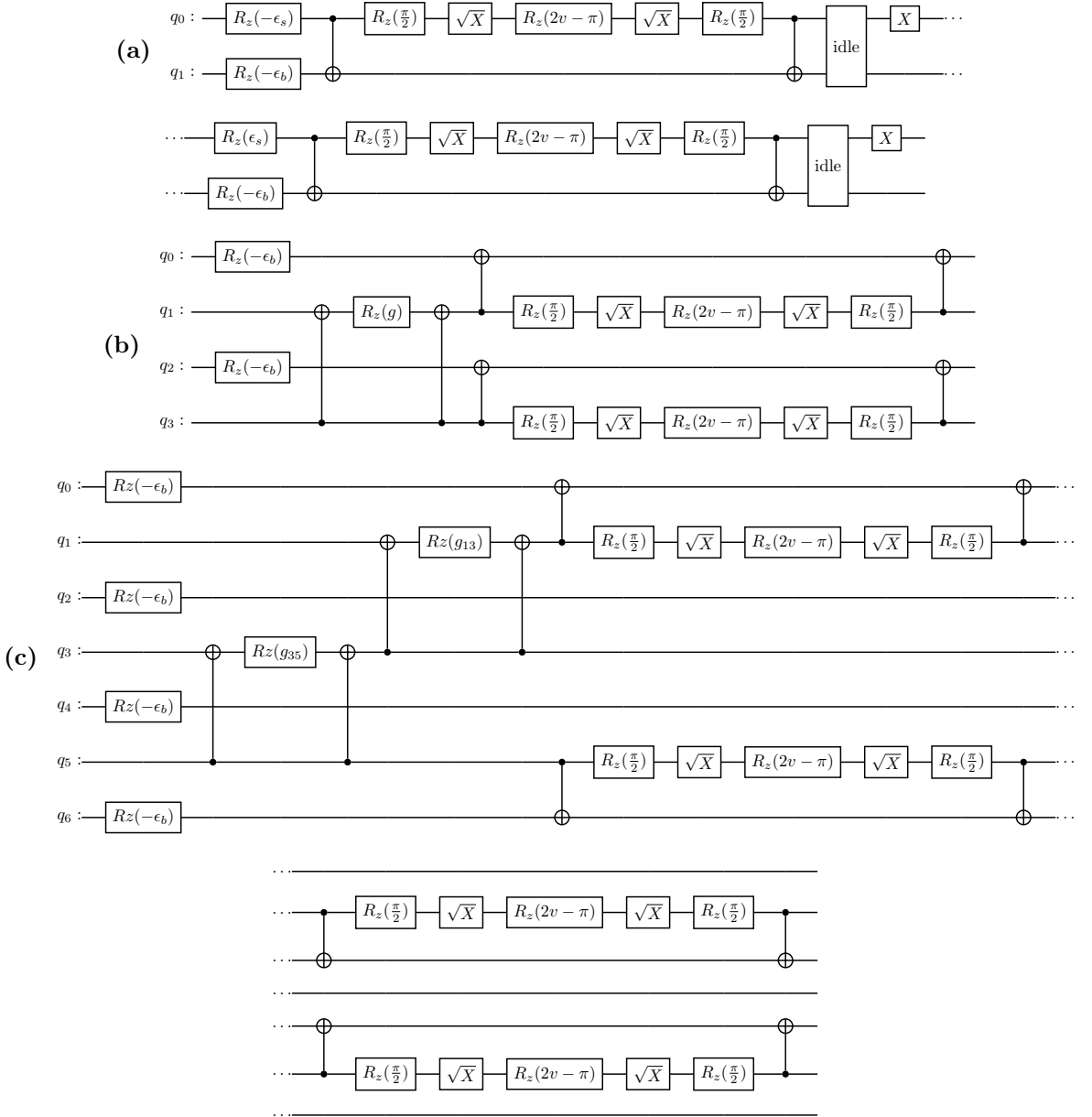
\begin{figure*}[htp]
\center
{\bf (a)}
\scalebox{0.7}{
\begin{quantikz}
\lstick{$q_0:$}     & \gate{R_z(-\epsilon_{s})}  & \ctrl{1}   & \gate{R_z(\frac{\pi}{2})}    & \gate{\sqrt{X}}   & \gate{R_z(2 v - \pi)}   & \gate{\sqrt{X}}   & \gate{R_z(\frac{\pi}{2})} & \ctrl{1}  & \gate[2]{\text{idle}}    & \gate{X} & \qw \dots \\
\lstick{$q_1:$}     & \gate{R_z(-\epsilon_{b})}  & \targ{}    & \qw                          & \qw               & \qw                     & \qw               & \qw                       & \targ{}   &                   & \qw      & \qw \dots \\
\end{quantikz}
}

\scalebox{0.7}{
\begin{quantikz}
\dots    & \gate{R_z(\epsilon_{s})}     & \ctrl{1}   & \gate{R_z(\frac{\pi}{2})}    & \gate{\sqrt{X}}   & \gate{R_z(2 v - \pi)}   & \gate{\sqrt{X}}   & \gate{R_z(\frac{\pi}{2})} & \ctrl{1}  & \gate[2]{\text{idle}}    & \gate{X} & \qw \\
\dots     & \gate{R_z(-\epsilon_{b})}      & \targ{}    & \qw                          & \qw               & \qw                     & \qw               & \qw                       & \targ{}   &                   & \qw      & \qw  \\
\end{quantikz}
}

{\bf (b)}
\scalebox{0.7}{
\begin{quantikz}
\lstick{$q_0:$} & \gate{R_z(-\epsilon_{b})}  & \qw       & \qw               & \qw       & \targ{}   & \qw               & \qw               & \qw                   & \qw              & \qw                         & \targ{}      & \qw   \\
\lstick{$q_1:$} & \qw               & \targ{}   & \gate{R_z(g)}  & \targ{}   & \ctrl{-1} & \gate{R_z(\frac{\pi}{2})} & \gate{\sqrt{X}}   & \gate{R_z(2 v - \pi)}   & \gate{\sqrt{X}}  & \gate{R_z(\frac{\pi}{2})}   & \ctrl{-1}    & \qw    \\
\lstick{$q_2:$} & \gate{R_z(-\epsilon_{b})}  & \qw         & \qw             & \qw       & \targ{}   & \qw               & \qw               & \qw                   & \qw              & \qw                         & \targ{}      & \qw      \\
\lstick{$q_3:$} & \qw               & \ctrl{-2} & \qw               & \ctrl{-2} & \ctrl{-1} & \gate{R_z(\frac{\pi}{2})} & \gate{\sqrt{X}}   & \gate{R_z(2 v - \pi)}   &  \gate{\sqrt{X}} & \gate{R_z(\frac{\pi}{2})}   & \ctrl{-1}    & \qw     \\
\end{quantikz}
}

{\bf (c)}
\scalebox{0.7}{
\begin{quantikz}
q_0: & \gate{Rz(-\epsilon_{b})}     & \qw       & \qw               & \qw       & \qw       & \qw               & \qw       & \targ{}   & \qw               & \qw               & \qw                   & \qw              & \qw                        & \targ{}   & \qw \dots \\
q_1: & \qw                          & \qw       & \qw               & \qw       & \targ{}   & \gate{Rz(g_{13})} & \targ{}   & \ctrl{-1} & \gate{R_z(\frac{\pi}{2})} & \gate{\sqrt{X}}   & \gate{R_z(2 v - \pi)} &  \gate{\sqrt{X}} & \gate{R_z(\frac{\pi}{2})}  & \ctrl{-1} & \qw \dots\\
q_2: & \gate{Rz(-\epsilon_{b})}     & \qw       & \qw               & \qw       & \qw       & \qw               & \qw       & \qw       & \qw               & \qw               & \qw                   & \qw              & \qw                        & \qw       & \qw \dots\\
q_3: & \qw                          & \targ{}   & \gate{Rz(g_{35})} & \targ{}   & \ctrl{-2} & \qw               & \ctrl{-2} & \qw       & \qw               & \qw               & \qw                   & \qw              & \qw                        & \qw       & \qw \dots\\
q_4: & \gate{Rz(-\epsilon_{b})}     & \qw       & \qw               & \qw       & \qw       & \qw               & \qw       & \qw       & \qw               & \qw               & \qw                   & \qw              & \qw                        & \qw       & \qw \dots\\
q_5: & \qw                          & \ctrl{-2} & \qw               & \ctrl{-2} & \qw       & \qw               & \qw       & \ctrl{1}  & \gate{R_z(\frac{\pi}{2})} & \gate{\sqrt{X}}   & \gate{R_z(2 v - \pi)} &  \gate{\sqrt{X}} & \gate{R_z(\frac{\pi}{2})}  & \ctrl{1}  & \qw \dots\\
q_6: & \gate{Rz(-\epsilon_{b})}     & \qw       & \qw               & \qw       & \qw       & \qw               & \qw       & \targ{}   & \qw               & \qw               & \qw                   & \qw              & \qw                        & \targ{}   & \qw \dots\\
\end{quantikz}
}
\scalebox{0.7}{
\begin{quantikz}
\dots & \qw         & \qw               & \qw               & \qw                   & \qw              & \qw                       & \qw       & \qw \\
\dots & \ctrl{1}    & \gate{R_z(\frac{\pi}{2})} & \gate{\sqrt{X}}   & \gate{R_z(2 v - \pi)} &  \gate{\sqrt{X}} & \gate{R_z(\frac{\pi}{2})}  & \ctrl{1}  & \qw \\
\dots & \targ{}     & \qw               & \qw               & \qw                   & \qw              & \qw                        & \targ{}   & \qw \\
\dots & \qw         & \qw               & \qw               & \qw                   & \qw              & \qw                        & \qw       & \qw \\
\dots & \targ{}     & \qw               & \qw               & \qw                   & \qw              & \qw                        & \targ{}   & \qw \\
\dots & \ctrl{-1}   & \gate{R_z(\frac{\pi}{2})} & \gate{\sqrt{X}}   & \gate{R_z(2 v - \pi)} &  \gate{\sqrt{X}} & \gate{R_z(\frac{\pi}{2})}  & \ctrl{-1} & \qw \\
\dots & \qw         & \qw               & \qw               & \qw                   & \qw              & \qw                        & \qw       & \qw \\
\end{quantikz}
}
\caption{Circuits that we run on IBM-Q.
(a) Two Trotter circuits corresponding to the Hamiltonian with one system spin (corresponding to qubit~$0$) and one auxiliary spin (qubit~$1$).
In the symmetrization, $X$-gates are added after the idling gates.
In (b) and (c), we only show the first Trotter circuit (the symmetrization is done analogously).
(b) The circuit corresponding to two system (qubits~$1$ and~$3$) and two auxiliary (qubits~$0$ and~$2$) spins.
(c) The circuit corresponding to three system (qubits~$1$,~$3$ and~$5$) and four auxiliary spins (qubits~$0$,~$2$,~$4$ and~$6$).}
\label{fig:circuits_all}
\end{figure*}

In this appendix, we give more detailed information on the circuits run on the IBM-Q devices.
The gate set of IBM-Q on the device we run is $\sqrt{X}$, $X$, CNOT, virtual $R_z$, and identity.

As outlined in the main text, symmetrization of the system qubit was achieved by adding $X$ gates to the circuits.
This procedure is depicted in Fig.~\ref{fig:circuits_all}(a),
which illustrates the symmetrized circuit for a single system qubit and one bath qubit.

For the cases of two and three system qubits, the corresponding circuits are presented in Fig.~\ref{fig:circuits_all}(b-c).
These figures focus on the Trotterized circuits, with the symmetrization process being analogous to that of the single system-bath qubit pair.

Additionally, we have made the circuit descriptions available in a serialized JSON-format 
compatible with the QoQo toolkit~\cite{qoqo, qoqo_files_online}.
The JSON files, being serializable, can be integrated into scripts for direct use
and should enhance the reproducibility of the simulation results.
The supplied QoQo files are compatible with various platforms,
including the QuEST~\cite{quest} numerical simulator, and can be executed on hardware through interfaces such as qoqo-qiskit or qoqo-for-braket,
offering flexibility in the choice of simulation execution environments.


\section{Numerical model}
\label{sec:appNumericalModel}
\begin{figure*}[hbt!]
\begin{center}
\includegraphics[width=2\columnwidth]{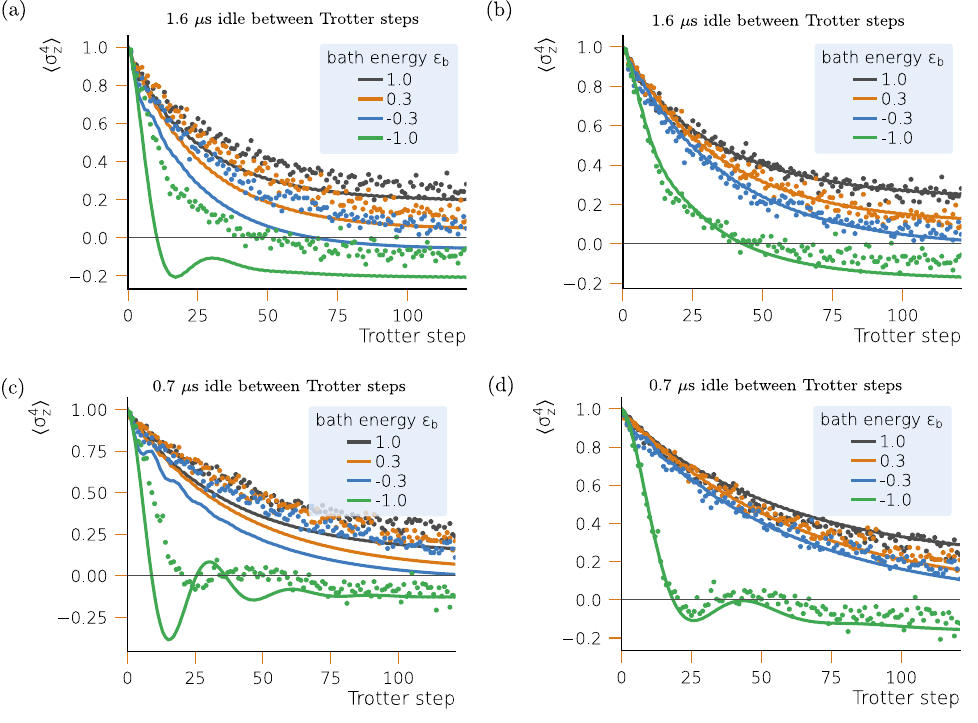}
\end{center}
\caption{
Time-evolution of a system-bath model consisting of one system spin coupled to on auxiliary spin.
The left column (a,c) compares the results obtained from IBM-Q (dots) to numerical simulation with calibration data obtained on the day the demonstration was conducted.
The right column (b,d) shows numerical simulations shown in Fig.~\ref{fig:1system_1bath}.
Here we included coherent errors by shifting parameters of the spin Hamiltonian and adapted the noise rates.
Notable in (c) is the oscillation period on the IBM-Q results, which is slower compared to the simulation, a clear sign of coherent error.
The simulation in (b) and (d) correspond to a coupling is $v=0.058$ in contrast to $v=0.1$ in (a) and (c).
Such effective change is consistent with the general magnitude and type of coherent errors observed in our additional analysis presented in Appendices~\ref{sec:noise_CNOT} and~\ref{sec:coherent_errors_to_effective_model}. 
Furthermore, in (b) and (d) additional $R_z$ rotations were introduced on the bath qubits, originating most probably in collected coherent Z-error during the idling phases (qubit frequency shift,
see Appendix~\ref{sec:noise_idling}).
This error vanishes in the system, due to the symmetrization algorithm.
The additional angles in (b) are specified as $0.31$ (case $\epsilon_b = 0.3$), $0.25$ ($\epsilon_b = -0.3$), $0.25$ ($\epsilon_b = -1.0$) and in (d)  $0.13$ (case $\epsilon_b = -1.0$).
Such effective change is consistent with a coherent error collected during additional qubit idling during the used~42~(a,b) or~20~(c,d)~idling gates
(the error being less than 0.01 per idling gate), see also Appendix~\ref{sec:noise_idling}.
If angles are not mentioned, we did not add $R_z$ rotations on bath qubits.
It should be mentioned that in several simulations (not presented in this manuscript) we have observed coherent Z-errors even larger than 1,
corresponding to coherent Z-error larger than 0.02 per idling gate.
The rate of damping, dephasing the depolarizing noise associated with CNOT gates are in (b,d) scaled down by a factor of 0.5.
}\label{fig:1system_1bath_appendix}
\end{figure*}
\begin{figure*}[hbt!]
\begin{center}
\includegraphics[width=2\columnwidth]{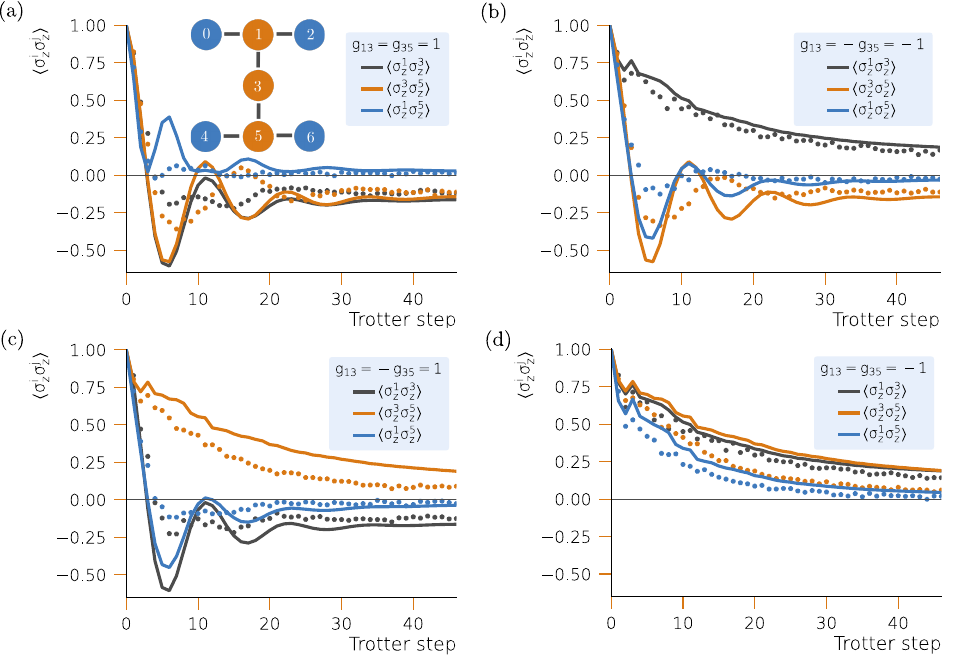}
\end{center}
\caption{
Time-evolution of a system-bath model consisting of three system and four bath qubits.
The figure compares the results obtained from IBM-Q (dots) to numerical simulations that use only the calibration data obtained
when the demonstration was conducted.
The numerical results have to be compared with the Fig.~\ref{fig:3system_4bath}, where
we have accounted for coherent errors by shifting parameters of the spin Hamiltonian, as described by Table~\ref{tab:parameters3s4b}.
}\label{fig:3system_4bath_appendix}
\end{figure*}
In this Appendix, we provide a description of the numerical model used to simulate the IBM-Q results.
The model is constructed based on quantum circuits and the corresponding calibration data, which have been recorded for all runs.

The model presented integrates both incoherent noise and coherent errors.
As described in the main text, the noise model includes spin depolarizing noise,
which affects all qubits while a gate is applied.
Additionally, during idle periods, the qubits are subject to both damping and dephasing processes.
To address coherent errors into our simulations, we have explored various approaches such as addition error-gates ($R_x(\theta)$, $R_y(\theta)$ and $R_z(\theta)$) after or before every gate.
Given the lack of precise knowledge regarding the nature and magnitude of coherent errors,
we have decided to adjust the parameters within the spin-Hamiltonian (rather than adding coherent errors to the circuits) when doing the fitting of the IBM-Q results.
This motivated by the fact that coherent gate errors lead to variations on the Hamiltonian, as discussed in Appendix~\ref{sec:coherent_errors_to_effective_model}.
The same approach is used for the rate of the incoherent noise, which can change after the calibration (we keep the form of the noise as described above).
The adaption of parameters is aimed at refining the numerical simulation to more closely mirror and understand the IBM-Q results.
Through this approach, we have achieved a satisfactory correlation between the numerical simulation outcomes and the IBM-Q demonstrations.

The shifting of Hamiltonian parameters and rescaling the noise rates differ for the three cases we consider in the main text.
In Fig.~\ref{fig:2system_2bath} we achieved a high degree of agreement using directly the calibration data and the noise rates from IBM-Q
with converting given gate errors to depolarizing noise as implied by universal formulas derived in Ref.~\cite{Abad2022}.
An exception here is the idling gate, which was assumed to be characterized by qubit damping and dephasing rates as implied by the $T_1$ and $T_2$ times
obtained from the calibration data, Table~\ref{tab:calibtration_data1}.
In Figs.~\ref{fig:1system_1bath} and~\ref{fig:3system_4bath}, it was necessary to adjust the parameters of the simulated spin-Hamiltonian
and the noise rates to better align with the IBM-Q results.
In Figs.~\ref{fig:1system_1bath_appendix} and~\ref{fig:3system_4bath_appendix}, we present the numerical simulations without these modifications.
A comparison showing the original and adapted parameters is given in Fig.~\ref{fig:1system_1bath_appendix} and in Table~\ref{tab:parameters3s4b}.
The magnitude and type of changes in the Hamiltonian parameters is very much consistent with the magnitude and type of observed coherent errors in noise characterization presented in Appendix~\ref{sec:NoiseAnalysis}.
\begin{table}[ht]
\begin{tabular}{c|c|c|c|c|c}
    & $g_{13}$  & $g_{35}$      & $v$           & noise scaling   & bath shift  \\
\hline
(a) & 0.6 (1)   & 0.6 (1)       & 0.16 (0.2)    & 1 (1)           & 0.55 (0)    \\
(b) & -1 (-1)   & 1 (1)         & 0.15 (0.2)    & 1.3 (1)         & 0.17 (0)    \\
(c) & 0.5 (1)   & -0.5 (-1)     & 0.17 (0.2)    & 1.4 (1)         & 0.17 (0)    \\
(d) & -1 (-1)   & -0.3 (-1)     & 0.15 (0.2)    & 1.3 (1)         & 0.07 (0)    \\
\hline
\end{tabular}
\centering
\caption{Parameters for the numerical fit of the three system- and four bath-qubit demonstration shown in Fig.~\ref{fig:3system_4bath}.
The parameters in bracket are the original parameters of the system-bath model.
The noise scaling gives the magnitude of the used noise in comparison to the calibration data.
The bath shift corresponds to the value of coherent Z-error collected over one Trotter circuit
(assumed here to be a constant for all bath qubits, to reduce the amount of fitting parameters).
Such effective change is consistent with the general magnitude and type of coherent errors of individual gates observed in Appendix~\ref{sec:NoiseAnalysis},
and with the relatively long depth of the circuits.
In particular, the large shift in the ZZ-interactions~$g$ may have contributions from a
direct ZZ-error during the application of the CNOT gates and
weak ZZ-interaction during idling~\cite{Sheldon_2016}.
}
\label{tab:parameters3s4b}
\end{table}

\end{document}